\documentclass{article} 
\usepackage{iclr2027_conference,times}

\usepackage{amsmath,amsfonts,bm}

\def\eqref#1{equation~\ref{#1}}

\def\1{\bm{1}}

\DeclareMathAlphabet{\mathsfit}{\encodingdefault}{\sfdefault}{m}{sl}
\SetMathAlphabet{\mathsfit}{bold}{\encodingdefault}{\sfdefault}{bx}{n}

\usepackage{hyperref}
\usepackage{url}

\usepackage[utf8]{inputenc} 
\usepackage[T1]{fontenc}    
\usepackage{booktabs}       
\usepackage{amsfonts}       
\usepackage{nicefrac}       
\usepackage{microtype}      
\usepackage{calc}
\usepackage{xcolor}         
\usepackage{adjustbox}

\usepackage{wrapfig}
\usepackage{microtype}
\usepackage{graphicx}
\usepackage{subcaption}
\usepackage{booktabs} 
\usepackage{adjustbox}

\usepackage{fancyhdr}
\usepackage{tocloft}
\usepackage{etoc}

\usepackage{algorithm}
\usepackage{algorithmic}

\usepackage{float}

\usepackage{amsmath}
\usepackage{amssymb}
\usepackage{mathtools}
\usepackage{amsthm}

\usepackage{multirow}
\usepackage{tabularx}

\usepackage[most]{tcolorbox}
\usepackage{xcolor,colortbl}
\tcbuselibrary{breakable,skins}

\usepackage{gradient-text}
\usepackage{xspace}

\usepackage{booktabs}
\usepackage{makecell}

\usepackage{pifont}

\usepackage{xltabular}
\usepackage{booktabs}
\usepackage{multirow}
\usepackage{makecell}
\usepackage{arydshln}  

\author{%
  \textbf{Xue Xia}\textsuperscript{1,5,*} \quad
  \textbf{Yue Wang}\textsuperscript{2,5,*} \quad
  \textbf{Shuizhou Chen}\textsuperscript{3,5,*} \quad
  \textbf{Lang Yu}\textsuperscript{5} \quad
  \textbf{Hao Wu}\textsuperscript{1,5} \\
  \textbf{Wenxuan Huang}\textsuperscript{1,5} \quad
  \textbf{Chengkai Yao}\textsuperscript{4,5} \quad
  \textbf{Zhangyang Gao}\textsuperscript{5,\dag} \quad
  \textbf{Siqi Sun}\textsuperscript{1,5} \\[2ex]
  \textsuperscript{1}Fudan University \quad
  \textsuperscript{2}University of Science and Technology of China \\
  \textsuperscript{3}Shanghai University \quad
  \textsuperscript{4}Shanghai Jiao Tong University \\
  \textsuperscript{5}Shanghai Artificial Intelligence Laboratory \\[2ex]
  \textsuperscript{*}Equal contribution \quad
  \textsuperscript{\dag}Project leader \\[2ex]
  \texttt{gaozhangyang@ailab.org.cn} \quad
  \texttt{siqisun@fudan.edu.cn}
}

\newcommand{\ours}{\gradientRGB{CellChain}{53,93,203}{10,10,80}\xspace}

\newcommand{\titleLogo}{%
    $\raisebox{-3.2mm}
    {\includegraphics[width=0.1\textwidth]{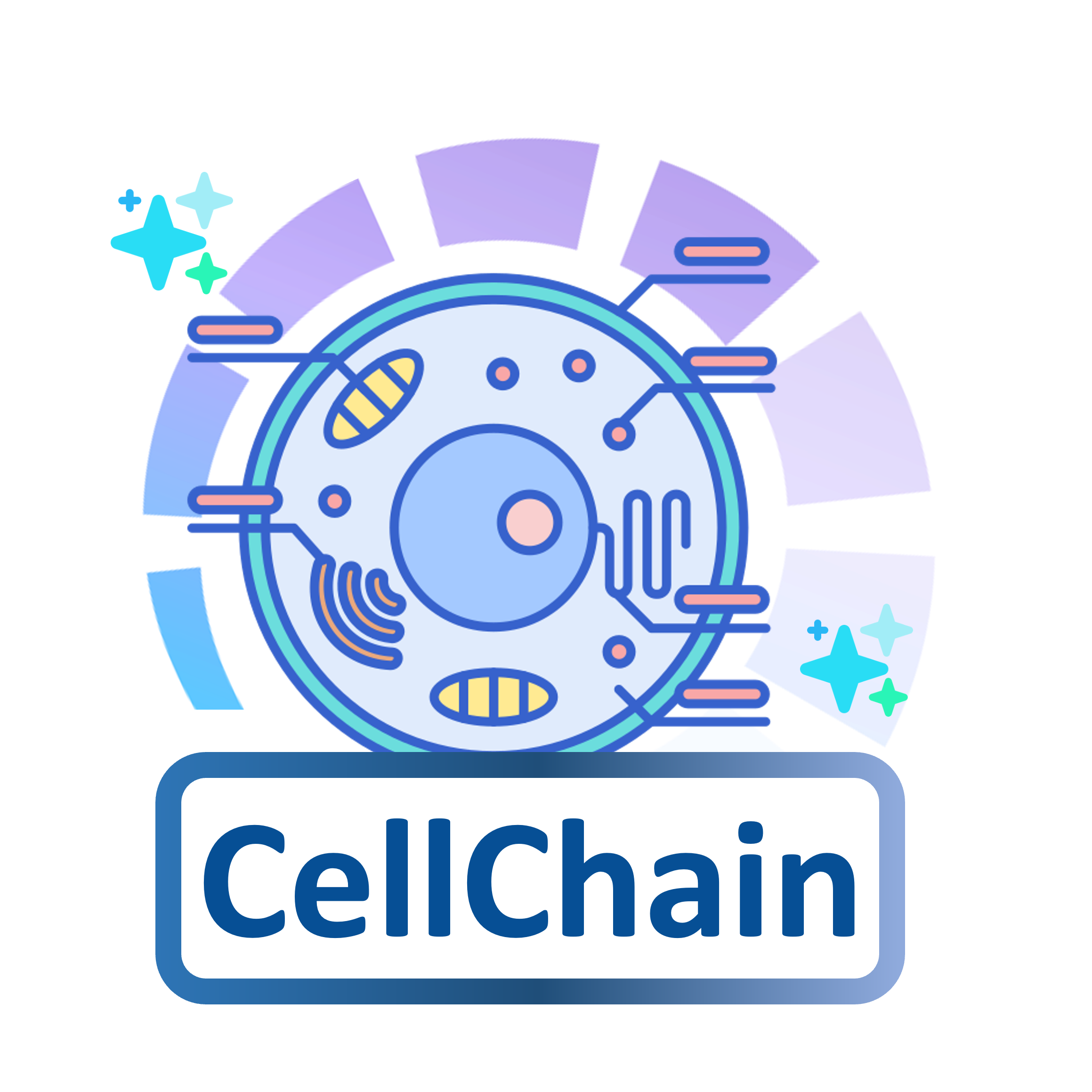}}$ 
}

\newcommand{\ncase}{590\xspace}

\newcommand{\ndataset}{17\xspace}

\newcommand{\nmodel}{18\xspace}

\title{\titleLogo \ours: Benchmarking Cross-Scale \\ Evidence Use for Perturbation Reasoning}

\iclrfinalcopy 
\begin{document}

\maketitle
\lhead{}

\begin{abstract}
Recent advances in foundation models and large-scale cellular perturbation datasets have driven growing interest in using LLMs to predict and reason about cellular responses.
Biological reasoning in this setting involves relating molecular mechanisms, cellular responses, and functional outcomes across scales.
Evaluating foundation models therefore requires assessing not only whether their predictions are correct, but also how those predictions depend on the biological evidence provided.
We introduce \ours{}, an experimentally grounded benchmark comprising 590 biological cases and 2,360 matched evidence-condition instances across gene-dependency prediction, cytokine-axis identification, and drug-sensitivity ranking.
\ours{} links perturbation experiments and external biological resources through a provenance-aware evidence graph spanning molecular, pathway, transcriptional, and phenotypic information.
Task-specific conditions vary evidence coverage, composition, or assignment while preserving questions, candidate answers, reference targets, and scoring rules.
Across 18 foundation models, the highest accuracies under prespecified target-evidence conditions are 57.5\%, 75.5\%, and 70.8\%, respectively, and broader evidence coverage does not consistently improve accuracy.
On the drug-sensitivity benchmark cohort, matched reference histories outperform swapped histories by an average of 13.0 percentage points in cluster-weighted accuracy.
Decision-transition analysis separates corrections, newly introduced errors, and wrong-answer switches, revealing substantial answer changes hidden by small net accuracy differences.
\ours{} provides controlled comparisons for attributing changes in model predictions to prespecified expansions of biological context, including changes in evidence coverage, composition, and biological assignment.
\end{abstract}

\section{Introduction}
\label{sec:introduction}
In recent years, researchers have developed a variety of benchmarks to evaluate large language models' biological reasoning capabilities, with an emphasis on cellular prediction tasks~\citep{zhang2025cellverse,zhao2026scarena} and pathway-centered mechanistic reasoning~\citep{zhao2025biomaze,kim2026pertreason}. While these benchmarks assess important aspects of biological understanding, they primarily evaluate whether models produce correct outcomes, providing limited insight into how those outcomes depend on the biological information supplied in context. This limitation is particularly important for cellular perturbation reasoning, where models must connect molecular interactions, pathway processes, transcriptional responses, and functional outcomes across biological scales~\citep{bunne2024virtualcell}. Existing studies further show that incorporating experimental observations or structured pathway knowledge does not necessarily lead to reliable predictions, particularly when models must interpret causal directionality or relate mechanistic information to downstream responses~\citep{wu2025perturbqa,kim2026pertreason, Yuan2026PlausibilityIN}. Evaluating biological reasoning therefore requires going beyond endpoint correctness to determine how model decisions change as specific biological evidence is introduced, combined, or reassigned under controlled conditions. 
\textit{Moreover, predictive improvements alone do not establish cross-scale integration, as models may exploit target-associated cues rather than integrate query-relevant experimental evidence.}

To address this gap, we introduce \ours, \textbf{an evidence-grounded benchmark for characterizing foundation model predictions under controlled cross-scale biological evidence.} \ours integrates large-scale perturbation experiments with external biological resources through an evidence graph connecting perturbations and cellular contexts to molecular annotations, pathway relations, transcriptional responses, and phenotypic outcomes. Reference answers are defined from experimental measurements or intervention metadata, and model-visible evidence retains its source provenance. Matched evaluation conditions preserve the question, candidate set, reference answer, and scoring rule while varying prespecified evidence-exposure configurations. This enables attribution of prediction changes to the corresponding context interventions while holding the underlying decision task fixed. This design spans genetic, cytokine, and chemical perturbations, supporting comparisons of predictive performance and evidence-dependent decisions without assuming that accuracy identifies the underlying reasoning mechanism. Table~\ref{tab:benchmark_comparison} situates \ours relative to prior benchmarks.

We evaluate \nmodel frontier foundation models, spanning proprietary and open-source models. The highest accuracies are 57.5\% for Gene Dependency Prediction, 75.5\% for Cytokine-Axis Inverse Recovery, and 70.8\% for Drug-Sensitivity Ranking, leaving substantial predictive headroom even under the target evidence conditions. Evidence-associated gains vary markedly across tasks and conditions, with broader context failing to yield consistent improvements. Moreover, substantial changes in model predictions can accompany nearly unchanged accuracy, revealing behavior obscured by aggregate performance alone. Model rankings also vary across tasks, with open-source models achieving competitive performance alongside proprietary systems. Together, these findings motivate evaluating both prediction accuracy and evidence-dependent decisions in biological reasoning.

\begingroup
\newcommand{\CSBFull}{\ding{52}}
\newcommand{\CSBPart}{partial}
\newcommand{\CSBNone}{\ding{56}}
\newcommand{\CSBBenchmark}[3]{%
  \makecell[l]{#1~(\citealp{#2}; \textit{#3})}%
}
\newcommand{\CSBPreprint}[2]{%
  \makecell[l]{#1~\citep{#2}}%
}
\begin{table*}[t]
\centering
\caption{Comparison between \ours and existing biological reasoning benchmarks.}
\label{tab:benchmark_comparison}
\scriptsize
\setlength{\tabcolsep}{3.3pt}
\renewcommand{\arraystretch}{1.15}

\newlength{\BenchColWidth}
\newlength{\ExpPertColWidth}
\newlength{\MultiIntColWidth}
\newlength{\CrossScaleColWidth}
\newlength{\ForwardColWidth}
\newlength{\ReverseColWidth}
\newlength{\RankingColWidth}
\newlength{\ProvenanceColWidth}
\newlength{\ControlledColWidth}
\newlength{\AttributionColWidth}
\setlength{\BenchColWidth}{5.2cm}
\setlength{\ExpPertColWidth}{1.2cm}
\setlength{\MultiIntColWidth}{1.3cm}
\setlength{\CrossScaleColWidth}{1.2cm}
\setlength{\ForwardColWidth}{1.2cm}
\setlength{\ReverseColWidth}{1.2cm}
\setlength{\RankingColWidth}{1.2cm}
\setlength{\ProvenanceColWidth}{1.2cm}
\setlength{\ControlledColWidth}{1.2cm}
\setlength{\AttributionColWidth}{1.2cm}

\resizebox{\textwidth}{!}{%
\begin{tabular}{@{}
>{\raggedright\arraybackslash}p{\BenchColWidth}
>{\centering\arraybackslash}p{\ExpPertColWidth}
>{\centering\arraybackslash}p{\MultiIntColWidth}
>{\centering\arraybackslash}p{\CrossScaleColWidth}
>{\centering\arraybackslash}p{\ForwardColWidth}
>{\centering\arraybackslash}p{\ReverseColWidth}
>{\centering\arraybackslash}p{\RankingColWidth}
>{\centering\arraybackslash}p{\ProvenanceColWidth}
>{\centering\arraybackslash}p{\ControlledColWidth}
>{\centering\arraybackslash}p{\AttributionColWidth}
@{}}
\toprule
& \multicolumn{3}{c}{\textbf{Biological grounding}}
& \multicolumn{3}{c}{\textbf{QA direction}}
& \multicolumn{3}{c}{\textbf{Evidence diagnostics}} \\
\cmidrule(lr){2-4}\cmidrule(lr){5-7}\cmidrule(lr){8-10}
\textbf{Benchmark} &
\makecell{Exp.\\perturb.} &
\makecell{Multiple\\interventions} &
\makecell{Joint\\cross-scale} &
\makecell{Forward\\effect} &
\makecell{Reverse\\cause} &
\makecell{Candidate\\ranking} &
\makecell{Linked\\provenance} &
\makecell{Controlled\\evidence} &
\makecell{Effect\\attribution} \\
\midrule
\CSBBenchmark{SOAR}{liu2025soar}{Briefings in Bioinformatics}
& \CSBNone & \CSBNone & \CSBNone & \CSBNone & \CSBNone & \CSBNone & \CSBNone & \CSBNone & \CSBNone \\
\CSBBenchmark{CellVerse}{zhang2025cellverse}{NeurIPS}
& \CSBFull & \CSBFull & \CSBPart & \CSBFull & \CSBNone & \CSBNone & \CSBNone & \CSBNone & \CSBNone \\
\CSBBenchmark{SC-Arena}{zhao2026scarena}{ICLR}
& \CSBFull & \CSBNone & \CSBPart & \CSBFull & \CSBNone & \CSBNone & \CSBPart & \CSBNone & \CSBNone \\
\CSBBenchmark{PerturbQA}{wu2025perturbqa}{ICLR}
& \CSBFull & \CSBNone & \CSBPart & \CSBFull & \CSBNone & \CSBNone & \CSBPart & \CSBPart & \CSBPart \\
\CSBBenchmark{VCWorld / GeneTAK}{wei2026vcworld}{ICLR}
& \CSBFull & \CSBNone & \CSBFull & \CSBFull & \CSBNone & \CSBNone & \CSBPart & \CSBPart & \CSBPart \\
\CSBPreprint{BioMaze}{zhao2025biomaze}
& \CSBPart & \CSBPart & \CSBPart & \CSBFull & \CSBNone & \CSBNone & \CSBFull & \CSBPart & \CSBPart \\
\CSBBenchmark{PertReasonQA}{kim2026pertreason}{ICML Workshop}
& \CSBFull & \CSBFull & \CSBFull & \CSBFull & \CSBNone & \CSBNone & \CSBPart & \CSBFull & \CSBFull \\
\CSBPreprint{AssayBench}{debrouwer2026assaybench}
& \CSBFull & \CSBNone & \CSBPart & \CSBFull & \CSBNone & \CSBFull & \CSBFull & \CSBNone & \CSBNone \\
\midrule
\textbf{\ours (Ours)}
& \CSBFull & \CSBFull & \CSBFull & \CSBFull & \CSBFull & \CSBFull & \CSBFull & \CSBFull & \CSBFull \\
\bottomrule
\end{tabular}%
}
\vspace{2pt}

\begin{minipage}{1\textwidth}
\centering
\footnotesize
\emph{Legend.} \CSBFull{} = explicit feature; \CSBPart{} = partial/indirect coverage; \CSBNone{} = not central to the design.
\end{minipage}
\end{table*}
\endgroup

\section{\ours}

\begin{figure}[h]
\centering
\includegraphics[width=1.0\linewidth]{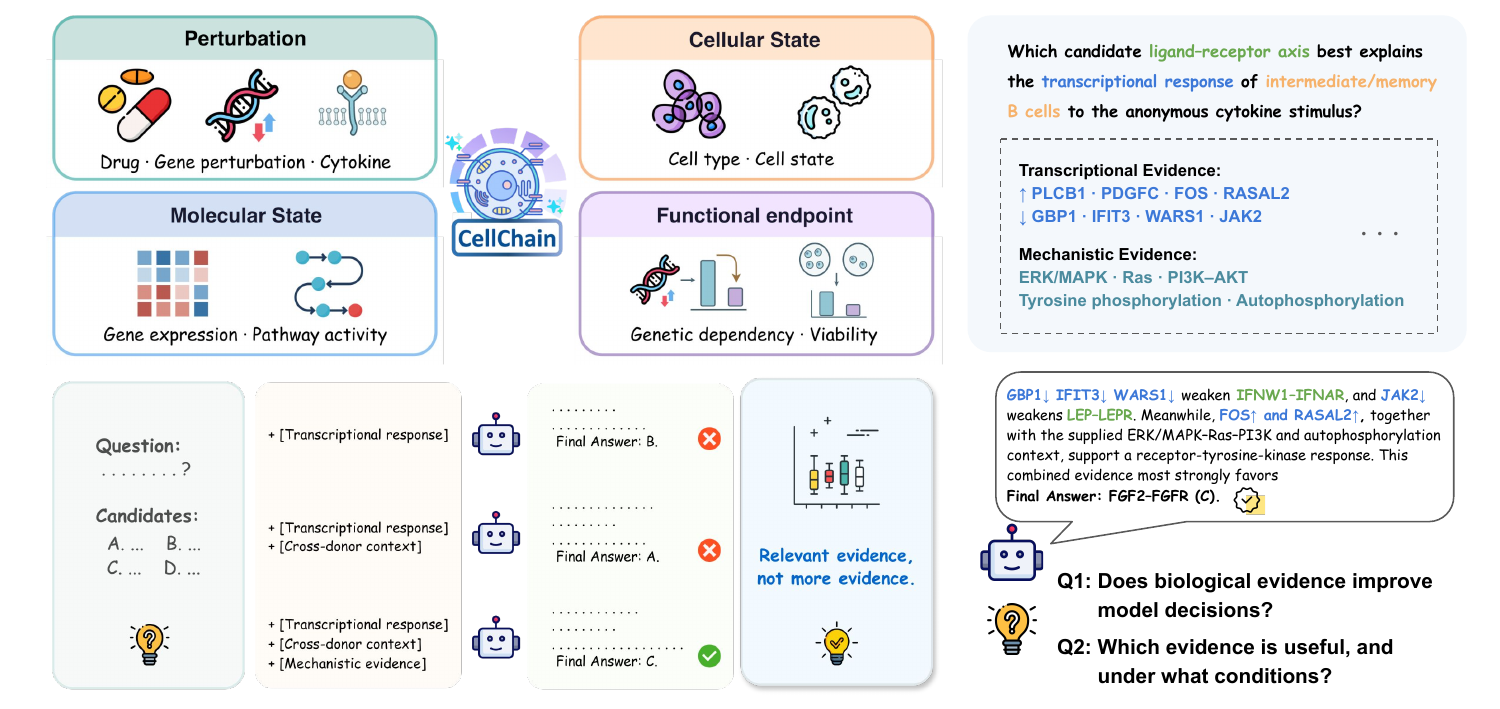}
\caption{
Overview of \ours.
\ours integrates cross-scale biological information into structured evidence and prediction targets for evidence-grounded QA construction.
The right panel illustrates a representative cytokine-axis evaluation with a held-out reference answer.
}
\label{fig:data_overview}
\end{figure}

We present \ours, a benchmark for evaluating foundation-model prediction behavior under controlled cross-scale biological evidence in cellular perturbation reasoning. \ours has three key features: (1) \textbf{Cross-scale evidence integration}, connecting molecular annotations, pathway relations, transcriptional responses, and phenotypic measurements through a traceable evidence graph (\S\ref{sec:evidence_integration}); (2) \textbf{Complementary perturbation reasoning tasks}, assessing decisions across genetic, cytokine, and chemical perturbations (\S\ref{sec:task_construction}); and (3) \textbf{Controlled evidence exposure}, examining how information content and biological associations affect model decisions under paired evaluation conditions (\S\ref{sec:evidence_exposure}). The following subsections describe data integration, task construction, evaluation protocols, quality control, and dataset characteristics.

\subsection{Cross-scale evidence integration}
\label{sec:evidence_integration}

We integrate three experimental anchor datasets spanning chemical, cytokine, and genetic perturbations: Tahoe-100M~\citep{zhang2025tahoe}, PBMC Parse10M~\citep{oesinghaus2025cytokine}, and Replogle Perturb-seq~\citep{replogle2022genotype}. These are linked to 14 external resources providing functional endpoints, molecular annotations, pathway knowledge, cellular context, and interaction evidence, including PRISM, DepMap, PubChem, ChEMBL, BindingDB, UniProtKB, Reactome, Gene Ontology/GOA, Cellosaurus, scBaseCount, CORUM, IntAct, SIGNOR, and Open Targets~\citep{corsello2020discovering,tsherniak2017dependency,kim2023pubchem,zdrazil2024chembl,gilson2016bindingdb,bateman2025uniprot,milacic2024reactome,geneontology2023,huntley2015goa,bairoch2018cellosaurus,scbasecount,tsitsiridis2023corum,deltoro2022intact,signor2023update,ochoa2023opentargets}. Table~\ref{tab:data_resources} summarizes their data types and roles in \ours.

All records are represented as provenance-tagged evidence units that preserve source-specific semantics. Target-defining endpoint fields are reserved for held-out query construction and scoring, whereas prespecified annotations, measurements, and eligible reference-case evidence may be exposed under task-specific conditions. Held-out query targets and all fields used to derive them are excluded from model inputs, and curated associations are not interpreted as direct causal or experimental effects unless explicitly supported by the source.

\paragraph{Entity and condition alignment.}
We align heterogeneous sources using shared biological entities and explicit experimental context. Entity mappings establish cross-source links, while metadata such as cell type or cell line, donor, dose, and duration determine whether linked evidence is applicable to a given case. This preserves source-specific measurement semantics and avoids treating biologically related records as experimentally equivalent.

\paragraph{Evidence graph organization.}
We organize cross-scale information into a provenance-aware evidence graph linking perturbations and cellular contexts to task-relevant evidence across biological levels. The evidence graph is a provenance-linked organization of record. For Gene Dependency, CRISPR perturbations are linked to UniProtKB annotations, Reactome pathways, and Perturb-seq transcriptional responses, while DepMap gene-effect measurements remain held out. For Cytokine-Axis Recovery, cytokine-treated cellular contexts are linked to measured transcriptional responses, process-level evidence, and system context, while the applied ligand--receptor axis remains held out. For Drug-Sensitivity Ranking, compound--cell-model contexts are linked to transcriptional responses and development-set reference histories, while the query drug-sensitivity endpoint remains held out.

\subsection{Benchmark Task Construction}
\label{sec:task_construction}

\ours defines three complementary perturbation prediction tasks. \textbf{Gene Dependency Prediction} predicts a category derived from held-out DepMap gene-effect scores, which quantify the effect of gene disruption on cell fitness. \textbf{Cytokine-Axis Inverse Recovery} identifies the applied cytokine through a ligand--receptor axis mapped from the recorded stimulus using frozen PBMC annotations. \textbf{Drug-Sensitivity Ranking} selects the more sensitive of two cell models exposed to the same compound, using held-out PRISM drug-response AUC as the reference endpoint.

\paragraph{Selecting Biological Cases.}
We construct eligible cases by linking perturbation measurements, cellular contexts, and task-specific reference targets. Gene cases require matched response and dependency records; Cytokine cases require treated cells with donor--cell-type-matched controls; Drug cases require complete transcriptional trajectories, QC-qualified sensitivity endpoints, and sufficient development-set reference-history coverage. These criteria ensure that the required model-visible evidence is available while reference targets remain independently defined. Additional case-selection procedures and evidence diagnostics are described in Appendix~\ref{app:evidence_attribution}.

\paragraph{Defining Prediction Targets.}
Gene Dependency uses three categories to distinguish strong, intermediate, and weak or absent dependency. Writing $g$ for the gene-effect score, we retain cases with $g\leq-1.05$, $-0.75\leq g\leq-0.55$, or $g\geq-0.25$, and exclude scores in the intervening intervals as boundary-ambiguous. The Cytokine reference is the recorded stimulus mapped to its ligand--receptor axis using the frozen PBMC annotation resource, rather than an axis inferred from the measured response. For Drug Sensitivity, lower PRISM AUC indicates greater sensitivity, and pairs require $|\Delta\mathrm{AUC}|\geq0.15$ to exclude near-tie comparisons.

\paragraph{Curating QA Tasks.} Each case contains a biological question, candidate answers, and a reference label. Gene presents the three dependency categories; Cytokine presents the target axis and three distractors, yielding a common four-choice inverse-identification task; Drug presents two cell models in both candidate orders, so that order-invariance can be assessed independently of task difficulty. Within a case, the question, candidate set, reference target, and scoring rule are held fixed across evidence conditions. This controls the task structure so that paired comparisons reflect the specified change in visible evidence or its assignment, rather than a different question or answer space.

\subsection{Controlled Evidence Exposure}
\label{sec:evidence_exposure}

We evaluate evidence dependence through controlled-evidence contrasts between matched biological contexts. Each base case retains the same question, candidate answers, held-out query target, and scoring rule, while the prespecified model-visible evidence or its biological assignment varies.

\paragraph{Controlling Information Coverage.}
We construct matched conditions that vary the biological evidence available for reasoning. Cumulative conditions progressively add evidence types and define incremental contrasts under a fixed order. Cytokine conditions present process annotations and system context separately and jointly under a shared evidence budget, while retaining the measured response in every condition. Baseline conditions retain the information required to specify each decision problem. The process annotations are linked to the recorded cytokine stimulus, and the joint condition reallocates the evidence budget between sources.

\paragraph{Testing Biological Associations.}
We further vary whether biological evidence is assigned to the entities it describes. Specifically, in Drug-Sensitivity Ranking, matched histories associate each candidate with development-set phenotype evidence from the same cell model, whereas swapped histories exchange these histories between candidates. To avoid leakage, reference histories are fixed before model evaluation from 36 disjoint development compounds after PRISM quality control (median fit $R^2\geq0.5$, replicate AUC range $\leq0.1$, positive slope, and ordered asymptotes); within-compound AUC comparisons determine sensitivity percentiles without using query-compound AUC or query transcriptional responses to select reference records. Both conditions retain the query responses and the same history-comparison instruction. Accordingly, their contrast measures sensitivity to the candidate--history assignment conditional on the supplied evidence in the benchmark cohort.

\paragraph{Separating Evidence from Targets.}
To preserve a clear separation between prediction evidence and scoring targets, held-out query endpoint values and explicit answer keys are excluded from model-visible inputs. Other measurements and reference-case evidence are exposed only when prespecified for the corresponding condition and remain subject to the same task-specific masking rules. In addition, task-specific identity masking is applied consistently across matched conditions, while Cytokine target-linked process annotations remain associated with the recorded stimulus. Finally, model-visible inputs are screened during quality control for prohibited endpoint fields, answer labels, direct target-revealing information, and condition-specific leakage cues.

\subsection{Dataset Analysis}
\label{sec:dataset_analysis}

Table~\ref{tab:dataset_statistics} summarizes the construction and coverage of \ours. The benchmark comprises \ncase problems derived from \ndataset biological resources spanning genetic, cytokine, and chemical perturbations, with evidence across molecular, pathway-process, transcriptional, and phenotypic levels. Experimental contexts include cell type, cell line, donor, dose, and duration where applicable. We apply four quality-control checks to each benchmark case. First, cross-source links must agree on the biological entity and cellular context, and unresolved mappings are excluded. Second, reference answers are recomputed from the underlying experimental endpoint or recorded intervention identity using the prespecified task rule; only cases with a unique valid answer are retained. Third, the question, candidate set, reference answer, and scoring rule remain fixed across matched evidence conditions, with only the specified evidence or associated instructions changing. Finally, model-visible inputs are screened for answer labels, prohibited query endpoint fields, and task-specific identity strings. Cases failing an applicable check in any condition are excluded as matched units. Six PhD researchers manually reviewed 150 sampled cases, of which 142 passed (94.7\%).

\begin{table}[t]
\centering
\caption{Construction statistics of \ours{}.}
\label{tab:dataset_statistics}
\newlength{\statsPanelGap}\setlength{\statsPanelGap}{0.4cm}
\newlength{\statsPanelWidth}\setlength{\statsPanelWidth}{0.47\linewidth}
\newlength{\statsLabelWidth}\setlength{\statsLabelWidth}{3.2cm}
\newlength{\statsGeneWidth}\setlength{\statsGeneWidth}{0.8cm}
\newlength{\statsCytokineWidth}\setlength{\statsCytokineWidth}{0.9cm}
\newlength{\statsDrugWidth}\setlength{\statsDrugWidth}{0.8cm}
\setlength{\tabcolsep}{0pt}
\renewcommand{\arraystretch}{1}
\begin{subtable}[t]{\statsPanelWidth}
\centering
\caption{Cases and coverage}
\label{tab:dataset_statistics_cases}
\footnotesize
\begin{tabular*}{\linewidth}{@{}>{\raggedright\arraybackslash}p{\statsLabelWidth}@{\extracolsep{\fill}}>{\raggedleft\arraybackslash}p{\statsGeneWidth}>{\raggedleft\arraybackslash}p{\statsCytokineWidth}>{\raggedleft\arraybackslash}p{\statsDrugWidth}@{}}
\toprule
Statistic & Gene & Cytokine & Drug \\
\midrule
Source records & 3,443 & 13,967 & 11,304 \\
Eligible cases & 2,092 & 267 & 737 \\
Benchmark cases & 240 & 110 & 240 \\
Benchmark / eligible (\%) & 11.5 & 41.2 & 32.6 \\
Perturbations & 219 & 11 & 26 \\
Cellular contexts & 2 & 6 & 34 \\
Donors & -- & 12 & -- \\
\bottomrule
\end{tabular*}
\end{subtable}\hspace{\statsPanelGap}%
\begin{subtable}[t]{\statsPanelWidth}
\centering
\caption{Evidence and evaluation}
\label{tab:dataset_statistics_evidence}
\footnotesize
\begin{tabular*}{\linewidth}{@{}>{\raggedright\arraybackslash}p{\statsLabelWidth}@{\extracolsep{\fill}}>{\raggedleft\arraybackslash}p{\statsGeneWidth}>{\raggedleft\arraybackslash}p{\statsCytokineWidth}>{\raggedleft\arraybackslash}p{\statsDrugWidth}@{}}
\toprule
Statistic & Gene & Cytokine & Drug \\
\midrule
Evidence units & 2,495 & 330 & 672 \\
Evidence assignments & 3,032 & 2,290 & 6,938 \\
Biological clusters & 219 & 11 & 26 \\
Evidence conditions & 4 & 4 & 4 \\
Task instances & 960 & 440 & 960 \\
Candidate orderings & 1 & 1 & 2 \\
Model requests & 960 & 440 & 1,920 \\
\bottomrule
\end{tabular*}
\end{subtable}
\end{table}

\section{Experiments}
\label{sec:experiments}


\subsection{Experiment Setup}
\label{sec:experimental_setup}

\paragraph{Evaluated Models.}
\label{sec:evaluated_models}

We evaluate the performance of state-of-the-art foundation models on \ours, spanning both proprietary and open-source systems with reasoning capabilities. The proprietary models evaluated are GPT-5.6 Sol~\citep{openai_gpt5_6_sol} and Gemini 3.8 Flash~\citep{gemini_3_8_flash}. Open-source models include DeepSeek-V4-Pro~\citep{deepseek_v4}, DeepSeek-V4.1-Flash~\citep{deepseek_v4_1_flash}, GLM-5.2~\citep{glm_5}, GLM-5.3~\citep{glm_5_3}, MiniMax-M3~\citep{minimax_m3}, Kimi-K3~\citep{kimi_k3}, Qwen3.8-Flash-Next~\citep{qwen_3_8_flash_next}, Qwen3.8-27B~\citep{qwen_3_8_27b}, Qwen3.6-35B-A3B~\citep{qwen_3_6}, GPT-OSS-20B~\citep{gpt_oss_20b}, Agents-A1~\citep{agents_a1}, Intern-S1-mini~\citep{intern_s1_mini}, Hunyuan-1.8B-Instruct~\citep{hunyuan_1_8b_instruct}, Hy3-FP8~\citep{hy3_fp8}, Llama-3.3-70B-Instruct~\citep{llama_3_3_70b}, and Llama-4-Scout-17B-16E-Instruct~\citep{llama_4_scout}.
This selection covers a range of model scales, providers, and reasoning configurations, accessed via standardized serving backends (API or vLLM). Detailed model specifications and versions are provided in \autoref{tab:model_configuration}.


\paragraph{Prompt Design.}
\label{sec:prompt_design}

We use a structured prompt consisting of task instructions, fixed query information, model-visible evidence, candidate answers, and a structured output schema, as shown in Figure~\ref{fig:controlled-prompt-template}. Questions, candidates, reference answers, and scoring remain fixed across matched conditions, while prespecified evidence content, its biological assignment, or associated instructions vary. Explicit answer keys and query-specific endpoint measurements are excluded from model-visible prompts.


\begin{figure}[t]
\centering

\resizebox{0.82\linewidth}{!}{%
\begin{minipage}{\linewidth}

\begin{tcolorbox}[
    enhanced,
    colback=white,
    colframe=black!40,
    boxrule=0.5pt,
    arc=2pt,
    left=5pt,
    right=5pt,
    top=4pt,
    bottom=5pt,
    title={Controlled Evidence Evaluation Template},
    colbacktitle=black!15,
    coltitle=black,
    fonttitle=\small\bfseries,
    fontupper=\footnotesize
]

\begin{tcolorbox}[
    colback=blue!4,
    colframe=blue!35,
    boxrule=0.4pt,
    arc=1.5pt,
    left=5pt,
    right=5pt,
    top=3pt,
    bottom=3pt,
    fontupper=\footnotesize
]
\textcolor{blue!55!black}{
\textbf{\textsf{SYSTEM}}
}\\[2pt]

You are a molecular and cellular biology expert. Answer the biological question using only the fixed task inputs and evidence visible under the current exposure condition. Treat annotations and biological associations as supporting evidence rather than direct measurements of the held-out query endpoint unless explicitly stated. Do not infer endpoint values that are not provided. Return exactly one valid JSON object conforming to the required output schema.

\end{tcolorbox}

\vspace{-2pt}

\begin{tcolorbox}[
    colback=black!3,
    colframe=black!35,
    boxrule=0.4pt,
    arc=1.5pt,
    left=5pt,
    right=5pt,
    top=3pt,
    bottom=3pt,
    fontupper=\footnotesize
]
\textcolor{black!65}{
\textbf{\textsf{MATCHED DECISION TASK --- FIXED ACROSS CONDITIONS}}
}\\[2pt]

\textbf{Question:}
\texttt{\{biological\_question\}}\\[2pt]

\textbf{Query and Fixed Observations:}
\texttt{\{query\_context\_and\_fixed\_observations\}}\\[2pt]

\textbf{Candidate Answers:}
\texttt{A: \{candidate\_A\}; B: \{candidate\_B\}; \ldots}

\end{tcolorbox}

\vspace{-2pt}

\begin{tcolorbox}[
    colback=purple!4,
    colframe=purple!40,
    boxrule=0.5pt,
    arc=1.5pt,
    left=5pt,
    right=5pt,
    top=3pt,
    bottom=3pt,
    fontupper=\footnotesize
]
\textcolor{purple!65!black}{
\textbf{\textsf{CONTROLLED EVIDENCE EXPOSURE --- VARIED BY CONDITION}}
}\\[2pt]

\textbf{Visible Cross-Scale Evidence:}\\[-1pt]

\texttt{[Molecular] \{molecular\_evidence\}}\\
\texttt{[Pathway/Process] \{pathway\_evidence\}}\\
\texttt{[Transcriptional] \{transcriptional\_evidence\}}\\
\texttt{[Reference-History] \{reference\_history\_evidence\}}\\[2pt]

\textit{Across matched evidence conditions, the visible evidence content or its biological assignment varies while the underlying decision task remains fixed.}

\end{tcolorbox}

\vspace{-2pt}

\begin{tcolorbox}[
    colback=black!3,
    colframe=black!35,
    boxrule=0.4pt,
    arc=1.5pt,
    left=5pt,
    right=5pt,
    top=3pt,
    bottom=3pt,
    fontupper=\footnotesize
]
\textcolor{black!65}{
\textbf{\textsf{OUTPUT CONTRACT --- FIXED ACROSS CONDITIONS}}
}\\[2pt]

\texttt{\{task\_specific\_output\_schema\}}

\end{tcolorbox}

\vspace{-2pt}

\begin{tcolorbox}[
    colback=yellow!8,
    colframe=orange!45,
    boxrule=0.4pt,
    arc=1.5pt,
    left=5pt,
    right=5pt,
    top=3pt,
    bottom=3pt,
    fontupper=\footnotesize
]
\textcolor{orange!65!black}{
\textbf{\textsf{BENCHMARK CONTROL --- NOT MODEL-VISIBLE}}
}\\[2pt]

\textbf{Held-out Query Target and Target-Defining Endpoint:}\\
\texttt{\{held\_out\_query\_target\}}

\end{tcolorbox}

\end{tcolorbox}

\end{minipage}%
}

\caption{Controlled-evidence evaluation template in \ours, with a fixed decision task and condition-specific evidence exposure.}
\label{fig:controlled-prompt-template}

\end{figure}


\paragraph{Implementation Details.}
\label{sec:implementation_detail}

Model predictions are evaluated against held-out query targets using task-specific metrics. Results are aggregated by model, task, and context arm to enable paired comparisons across conditions. We set the sampling temperature to 0.0 to reduce variability. The maximum output length is set to 8,192 tokens, with an increased budget of 32,768 tokens for reasoning-enabled evaluations. Each prompt is constructed with up to 15 evidence units, using a prespecified evidence selection and ordering procedure applied consistently across models.

\subsection{Evaluation Metrics}
\label{sec:evaluation_metrics}

We report strict accuracy together with macro-averaged precision, recall, and F1, all normalized to $[0,1]$. Let $i$ index biological cases and $o$ index candidate-order presentations: $O{=}1$ for Gene Dependency and Cytokine Axis, and $O{=}2$ (reversed candidate order) for Drug Sensitivity. Let $\hat{y}_{m,i,o}^{(a)}$ be model $m$'s prediction under evidence condition $a$, and $y_{i,o}$ the reference answer for that presentation. We score each presentation and average within case:
\[
s_{m,i,o}^{(a)}=\mathbb{I}\!\left[\hat{y}_{m,i,o}^{(a)}=y_{i,o}\right],
\qquad
\bar{s}_{m,i}^{(a)}=\frac{1}{O}\sum_{o=1}^{O}s_{m,i,o}^{(a)}.
\]
For Drug Sensitivity, $\bar{s}_{m,i}^{(a)}\in\{0,\tfrac12,1\}$ because the two candidate-order presentations encode the same biological target with reversed answer positions.

\paragraph{Endpoint Performance.}
Endpoint accuracy averages $\bar{s}_{m,i}^{(a)}$ equally across the $N$ cases:
\[
\mathrm{Acc}_{m}(a)=\frac{1}{N}\sum_{i=1}^{N}\bar{s}_{m,i}^{(a)}.
\]
For Drug, accuracy first averages the two candidate-order presentations within each case and then averages across cases. Macro precision, recall, and F1 use pooled presentation-level counts, retaining invalid outputs as false negatives. Cytokine and Drug macro metrics describe candidate-position labels. Endpoint accuracy is case-weighted, whereas controlled contrasts are cluster-weighted.

\paragraph{Controlled-Evidence Contrasts.}
Cases are grouped into biological clusters $\mathcal{C}_1,\ldots,\mathcal{C}_K$ by perturbed gene, cytokine axis, or compound, and contrasts weight clusters equally. For conditions $a,b$, the within-cluster contrast is
\[
\delta_{m,k}(a,b)=\frac{1}{|\mathcal{C}_k|}\sum_{i\in\mathcal{C}_k}
\left(\bar{s}_{m,i}^{(a)}-\bar{s}_{m,i}^{(b)}\right),
\]
and the model- and benchmark-level contrasts give equal weight to clusters and models:
\[
\Delta_{m,\mathrm{Acc}}(a,b)=\frac{1}{K}\sum_{k=1}^{K}\delta_{m,k}(a,b),
\qquad
\overline{\Delta}_{\mathrm{Acc}}(a,b)=\frac{1}{M}\sum_{m=1}^{M}\Delta_{m,\mathrm{Acc}}(a,b).
\]
Positive values indicate higher accuracy under $a$. For Drug, $\Delta_{\mathrm{Combined}}$ denotes Matched minus Background, since this comparison jointly adds responses, histories, and history-comparison instructions. For Cytokine, we define
\[
\Gamma =
A_{\mathrm{Joint}}-A_{\mathrm{Process}}-A_{\mathrm{System}}+A_{\mathrm{Response}},
\]
where each $A$ is cluster-weighted. Because the Joint condition reallocates a fixed evidence budget across Process and System evidence, $\Gamma$ is interpreted as a joint-condition deviation from additivity rather than a factorial interaction effect. Negative $\Gamma$ indicates that Joint performance falls below the additive expectation defined by the separate conditions, potentially reflecting both evidence combination and budget-induced composition changes.

\paragraph{Decision Transitions.}
\label{sec:decision_transitions}
For a transition from $b$ to $a$, paired within case and presentation, let $p_b,p_a$ be normalized outputs (invalid outputs share a common state). Under the same cluster and model weights as above:
\[
\begin{aligned}
F&=\Pr\nolimits_w(p_a\ne p_b), &
C&=\Pr\nolimits_w(p_b\ne y,\,p_a=y), \\
D&=\Pr\nolimits_w(p_b=y,\,p_a\ne y), &
W&=\Pr\nolimits_w(p_a\ne p_b,\,p_b\ne y,\,p_a\ne y).
\end{aligned}
\]
It follows that $F = C + D + W$ and $\overline{\Delta}_{\mathrm{Acc}} = C - D$, where $W$ captures both switches between incorrect answers and transitions involving invalid outputs.

\subsection{Main Results}
\label{sec:main_results}

\paragraph{Cross-scale biological reasoning remains challenging for current foundation models.}
Across evaluated foundation models, the highest accuracies under the prespecified target-evidence conditions are 57.5\% for Gene Dependency, 75.5\% for Cytokine Axis, and 70.8\% for Drug Sensitivity (Table~\ref{tab:main_results}a). The corresponding error rates under the task-specific scoring rules are 42.5\%, 24.5\%, and 29.2\%. The three tasks have different candidate counts and should not be ranked by raw accuracy alone; their uniform-choice levels are 33.3\%, 25.0\%, and 50.0\%. Within each task, accuracy spans 32.5--57.5\%, 25.5--75.5\%, and 33.3--70.8\%, showing that the observed headroom is accompanied by substantial differences in model capability rather than a single shared performance ceiling. In particular, several models remain near uniform-choice on Gene Dependency despite receiving the full cumulative evidence, whereas stronger results on Cytokine Axis and Drug Sensitivity show the corresponding decision problems contain usable predictive information. Together, these results establish a challenging evaluation across distinct perturbation modalities while motivating the condition-level analysis of how models respond to supplied evidence.

\begin{table*}[t]
  \centering
  \caption{Main results in \ours{}(\%).}
  \vspace{-3pt}
  \label{tab:main_results}
  \scriptsize
  \setlength{\tabcolsep}{2pt}
  \renewcommand{\arraystretch}{1.1}

  \newcolumntype{A}{>{\raggedright\arraybackslash}p{3.4cm}}
  \newcolumntype{G}{>{\centering\arraybackslash}p{0.7cm}}
  \newcolumntype{C}{>{\centering\arraybackslash}p{0.7cm}}
  \newcolumntype{R}{>{\centering\arraybackslash}p{0.78cm}}
  \newcolumntype{B}{>{\raggedright\arraybackslash}p{3.4cm}}
  \newcolumntype{E}{>{\centering\arraybackslash}p{0.98cm}}
  \newcolumntype{F}{>{\centering\arraybackslash}p{0.98cm}}
  \newcolumntype{H}{>{\centering\arraybackslash}p{1.1cm}}

  \textbf{(a) Endpoint performance under target evidence conditions}\\[2pt]
  \resizebox{\textwidth}{!}{%
  \begin{tabular}{A|GGGG|CCCC|RRRR}
      \noalign{\hrule height 0.8pt}
      \textbf{Model}
      & \multicolumn{4}{c|}{\textbf{Gene Dependency}}
      & \multicolumn{4}{c|}{\textbf{Cytokine Axis}}
      & \multicolumn{4}{c}{\textbf{Drug Sensitivity}} \\
      & Prec. & Rec. & F1 & Acc. & Prec. & Rec. & F1 & Acc. & Prec. & Rec. & F1 & Acc. \\
      \hline
      \multicolumn{13}{>{\columncolor{gray!15}}c}{\rule[-2pt]{0pt}{11pt}\textit{Open-Source Models}} \\
      \hline
      DeepSeek-V4-Pro & 51.5 & 42.1 & 32.2 & 42.1 & 68.8 & 67.2 & 67.0 & 66.4 & 69.8 & 69.8 & 69.8 & 69.8 \\
      GLM-5.2 & 50.2 & 46.7 & 37.9 & 46.7 & 54.8 & 55.2 & 53.6 & 53.6 & 63.2 & 62.9 & 62.7 & 62.9 \\
      MiniMax-M3 & 38.8 & 44.2 & 37.6 & 44.2 & 65.4 & 65.8 & 65.3 & 65.5 & 67.7 & 67.7 & 67.7 & 67.7 \\
      Kimi-K3 & 52.3 & 54.2 & 48.0 & 54.2 & 70.7 & 70.4 & 70.3 & 70.0 & 70.0 & 70.0 & 70.0 & 70.0 \\
      DeepSeek-V4.1-Flash & 47.5 & 52.9 & 43.4 & 52.9 & 61.1 & 61.6 & 61.1 & 60.9 & 43.7 & 43.8 & 43.7 & 43.8 \\
      GLM-5.3 & 53.9 & 55.0 & 47.4 & 55.0 & 71.4 & 71.6 & 71.1 & 70.9 & 68.7 & 68.5 & 68.6 & 68.5 \\
      Qwen3.6-35B-A3B & 44.6 & 34.6 & 19.2 & 34.6 & 60.2 & 59.9 & 59.9 & 60.0 & 67.9 & 67.9 & 67.9 & 67.9 \\
      Qwen3.8-27B & 44.5 & 34.2 & 18.4 & 34.2 & 59.1 & 59.7 & 59.2 & 59.1 & 70.6 & 70.6 & 70.6 & 70.6 \\
      GPT-OSS-20B & 11.1 & 33.3 & 16.7 & 33.3 & 38.3 & 38.9 & 38.0 & 38.2 & 70.8 & 70.8 & 70.8 & 70.8 \\
      Agents-A1 & 39.0 & 35.0 & 20.6 & 35.0 & 45.8 & 45.8 & 45.3 & 45.5 & 46.0 & 46.0 & 45.8 & 46.0 \\
      Intern-S1-mini & 16.7 & 32.9 & 17.3 & 32.9 & 42.0 & 40.6 & 40.7 & 40.0 & 44.1 & 40.4 & 42.1 & 40.4 \\
      Hunyuan-1.8B-Instruct & 11.3 & 32.5 & 16.7 & 32.5 & 31.1 & 26.7 & 12.9 & 25.5 & 25.0 & 50.0 & 33.3 & 50.0 \\
      Qwen3.8-Flash-Next & 34.3 & 47.9 & 39.1 & 47.9 & 79.7 & 75.2 & 76.8 & 74.5 & 69.7 & 69.6 & 69.7 & 69.6 \\
      Llama-3.3-70B-Instruct & 11.1 & 33.3 & 16.7 & 33.3 & 39.2 & 33.7 & 29.9 & 32.7 & 33.2 & 33.3 & 33.2 & 33.3 \\
      Llama-4-Scout-17B-16E-Instruct & 11.1 & 33.3 & 16.7 & 33.3 & 29.2 & 35.7 & 30.0 & 36.4 & 35.4 & 35.4 & 35.4 & 35.4 \\
      Hy3-FP8 & 11.1 & 33.3 & 16.7 & 33.3 & 60.8 & 61.2 & 60.4 & 60.9 & 60.6 & 55.4 & 49.2 & 55.4 \\
      \hline
      \multicolumn{13}{>{\columncolor{gray!15}}c}{\rule[-2pt]{0pt}{11pt}\textit{Proprietary Models}} \\
      \hline
      GPT-5.6-Sol & 52.9 & 57.5 & 52.2 & 57.5 & 72.1 & 72.4 & 71.9 & 71.8 & 68.8 & 68.8 & 68.7 & 68.8 \\
      Gemini-3.8-Flash & 68.9 & 53.3 & 43.0 & 53.3 & 75.7 & 76.3 & 75.7 & 75.5 & 70.2 & 70.2 & 70.2 & 70.2 \\
      \noalign{\hrule height 0.8pt}
  \end{tabular}
  }%

  \vspace{2pt}
  \textbf{(b) Controlled evidence contrasts}\\[2pt]
  \resizebox{\textwidth}{!}{%
  \begin{tabular}{B|EEE|FFF|HHH}
      \noalign{\hrule height 0.8pt}
      \textbf{Model}
      & \multicolumn{3}{c|}{\textbf{Gene Dependency}}
      & \multicolumn{3}{c|}{\textbf{Cytokine Axis}}
      & \multicolumn{3}{c}{\textbf{Drug Sensitivity}} \\
      & \shortstack{$\Delta$Mol.} & \shortstack{$\Delta$Path.} & \shortstack{$\Delta$Trans.}
      & \shortstack{$\Delta$Process} & \shortstack{$\Delta$System} & $\Gamma$
      & \shortstack{$\Delta$Response} & \shortstack{$\Delta$Combined} & \shortstack{$\Delta$Assign.} \\
      \hline
      \multicolumn{10}{>{\columncolor{gray!15}}c}{\rule[-2pt]{0pt}{11pt}\textit{Open-Source Models}} \\
      \hline
      DeepSeek-V4-Pro & +0.5 & +8.2 & $-$1.1 & +41.8 & +0.9 & +0.0 & +2.1 & +12.6 & +27.0 \\
      GLM-5.2 & +0.0 & +11.9 & $-$0.5 & +44.5 & +0.0 & $-$6.4 & $-$6.9 & +8.9 & +19.7 \\
      MiniMax-M3 & +0.5 & +10.3 & $-$1.8 & +43.6 & $-$6.4 & +10.0 & $-$6.2 & +11.1 & +25.1 \\
      Kimi-K3 & +17.1 & $-$3.4 & +1.1 & +60.9 & +7.3 & $-$15.5 & +0.2 & +22.1 & +27.0 \\
      DeepSeek-V4.1-Flash & +5.7 & +10.3 & $-$0.7 & +39.1 & +3.6 & $-$11.8 & $-$9.6 & $-$7.3 & $-$5.3 \\
      GLM-5.3 & +13.5 & +3.4 & +0.5 & +54.5 & +4.5 & $-$8.2 & $-$5.6 & +10.7 & +25.9 \\
      Qwen3.6-35B-A3B & +0.0 & +0.5 & +0.5 & +46.4 & $-$1.8 & +0.0 & +3.8 & +12.1 & +23.9 \\
      Qwen3.8-27B & +0.0 & +0.0 & +0.7 & +53.6 & $-$0.9 & $-$4.5 & $-$1.0 & +16.1 & +28.2 \\
      GPT-OSS-20B & +0.0 & +0.0 & +0.0 & +21.8 & $-$4.5 & +1.8 & +1.8 & +14.7 & +28.3 \\
      Agents-A1 & +0.0 & +1.4 & +0.2 & +40.9 & $-$2.7 & $-$10.0 & $-$2.8 & $-$2.4 & $-$0.9 \\
      Intern-S1-mini & +0.0 & +0.5 & $-$0.9 & +40.9 & +0.9 & $-$9.1 & $-$4.0 & $-$6.8 & $-$11.8 \\
      Hunyuan-1.8B-Instruct & +0.0 & +0.5 & $-$1.1 & +0.0 & +0.0 & +0.9 & +0.0 & +0.0 & +0.0 \\
      Qwen3.8-Flash-Next & +0.0 & +15.5 & $-$5.0 & +56.4 & $-$4.5 & $-$2.7 & $-$2.4 & +12.2 & +26.1 \\
      Llama-3.3-70B-Instruct & +0.0 & +0.0 & +0.0 & +27.3 & $-$2.7 & $-$3.6 & +0.1 & $-$10.8 & $-$20.6 \\
      Llama-4-Scout-17B-16E-Instruct & +0.0 & +0.0 & +0.0 & +31.8 & $-$0.9 & $-$10.9 & +0.3 & $-$7.6 & $-$16.5 \\
      Hy3-FP8 & +0.0 & +0.0 & +0.0 & +57.3 & +0.9 & $-$8.2 & $-$7.9 & +0.7 & +5.0 \\
      \hline
      \multicolumn{10}{>{\columncolor{gray!15}}c}{\rule[-2pt]{0pt}{11pt}\textit{Proprietary Models}} \\
      \hline
      GPT-5.6-Sol & +4.8 & +13.5 & +0.0 & +50.9 & $-$0.9 & $-$6.4 & $-$8.9 & +10.6 & +25.7 \\
      Gemini-3.8-Flash & +18.0 & $-$4.3 & +0.5 & +50.9 & +1.8 & $-$7.3 & $-$7.2 & +13.7 & +26.4 \\
      \noalign{\hrule height 0.8pt}
  \end{tabular}
  }%
  \vspace{3pt}

\begin{minipage}{1\textwidth}
\scriptsize
\emph{Notes.}
Panel (a) evaluates full cumulative evidence for Gene Dependency, joint evidence for Cytokine Axis, and the matched condition for Drug Sensitivity.
Panel (b) reports: Gene, $\Delta_{\mathrm{Mol.}}=\mathrm{Molecular}-\mathrm{Base}$, $\Delta_{\mathrm{Path.}}=\mathrm{Pathway}-\mathrm{Molecular}$, and $\Delta_{\mathrm{Trans.}}=\mathrm{Transcriptional}-\mathrm{Pathway}$; Cytokine, $\Delta_{\mathrm{Process}}=\mathrm{Process}-\mathrm{Response}$, $\Delta_{\mathrm{System}}=\mathrm{System}-\mathrm{Response}$, and $\Gamma=\mathrm{Joint}-\mathrm{Process}-\mathrm{System}+\mathrm{Response}$, interpreted as a joint-condition deviation from additivity under the shared evidence budget; Drug, $\Delta_{\mathrm{Response}}=\mathrm{QueryResponse}-\mathrm{Background}$, $\Delta_{\mathrm{Combined}}=\mathrm{Matched}-\mathrm{Background}$, and $\Delta_{\mathrm{Assign.}}=\mathrm{Matched}-\mathrm{Swapped}$.
\end{minipage}
\end{table*}

\paragraph{Evidence attribution reveals heterogeneous effects across scales and assignments.}
Paired, cluster-weighted contrasts show that broader context does not uniformly improve predictions (Table~\ref{tab:main_results}b; Appendix~\ref{app:bootstrap_CI}). Across 18 models, molecular and pathway evidence improve Gene Dependency accuracy by 3.3 and 3.8 points, while the transcriptional increment is $-0.4$ points (95\% CI [$-1.1$, $+0.3$]). For Cytokine Axis, process-associated evidence yields a $+42.4$-point gain, while system context contributes $-0.3$ points and the joint-condition deviation is $-5.1$ points (95\% CI [$-8.9$, $-1.6$]). Because the process annotations are linked to the target stimulus, this gain reflects the predictive value of target-linked process context, with response--mechanism integration unresolved. For Drug Sensitivity, matched histories outperform swapped histories by 13.0 points (95\% CI [$+0.5$, $+24.5$]), whereas Matched versus Background is smaller and uncertain ($+6.2$ points; 95\% CI [$-0.3$, $+12.2$]). Overall, context expansion effects depend on evidence type, composition, and biological assignment, motivating controlled attribution beyond endpoint accuracy.

\paragraph{The performance gap between proprietary and open-source foundation models varies by biological task.}
Proprietary models achieve the highest observed accuracy on Gene Dependency and Cytokine Axis, whereas open-source models lead on Drug Sensitivity. On Gene Dependency, GPT-5.6-Sol reaches 57.5\%, compared with 55.0\% for the strongest open-source model, GLM-5.3. On Cytokine Axis, Gemini-3.8-Flash obtains 75.5\%, followed closely by Qwen3.8-Flash-Next at 74.5\%. The ordering reverses on Drug Sensitivity, where GPT-OSS-20B and Qwen3.8-27B achieve 70.8\% and 70.6\%, respectively, compared with 70.2\% for Gemini-3.8-Flash. These differences are descriptive rankings rather than evidence of statistically significant differences between model groups. More broadly, the variation in task leaders indicates that performance differences between proprietary and open-source systems are task dependent rather than consistent across \ours{}. These reversals highlight the value of \ours{}' complementary task design in revealing task-specific capability differences obscured by aggregate model rankings.

\begin{figure}[t]
  \centering
  \includegraphics[width=1\textwidth]{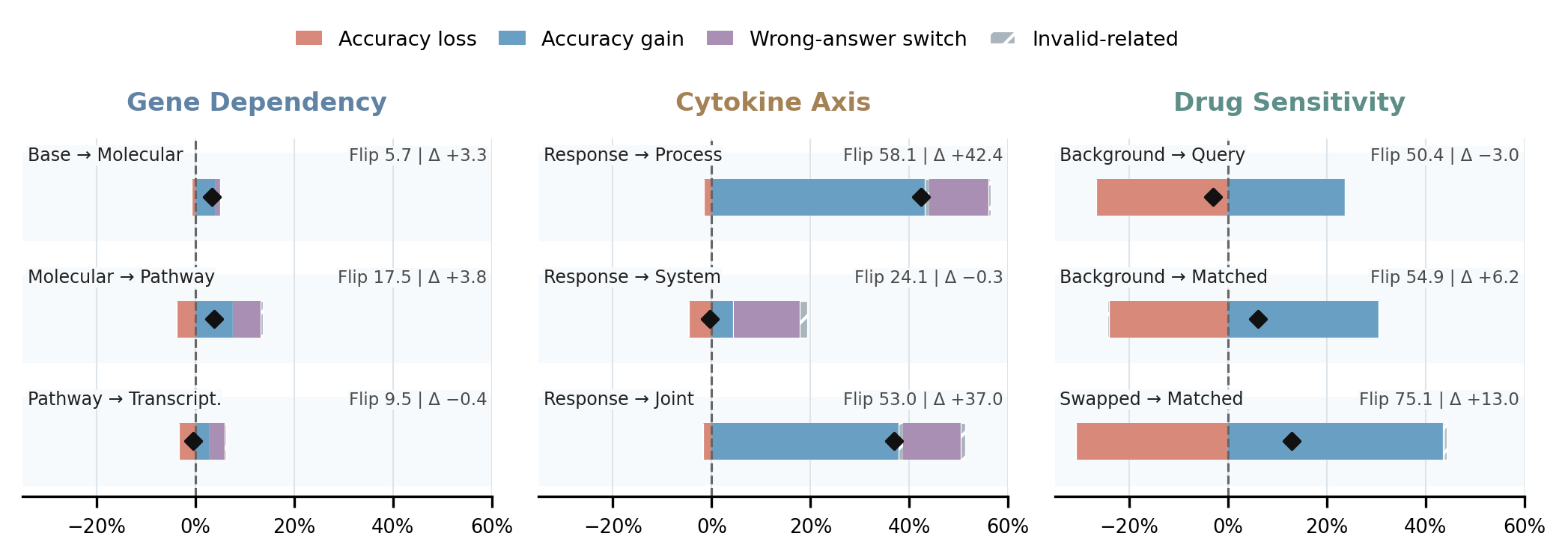}
  \caption{
  Decision transitions across evidence conditions. Bars show losses (leftward), gains, wrong-answer switches, and invalid-related changes; diamonds mark net $\Delta$ (pp), with flip rates annotated. 
  }
  \label{fig:decision_transitions}
\end{figure}

\paragraph{Small net accuracy changes conceal substantial decision turnover.}
Figure~\ref{fig:decision_transitions} decomposes prediction changes into corrections, newly introduced errors, wrong-answer switches, and invalid-related transitions. Gene Pathway-to-Transcriptional changes 9.5\% of predictions for a $-0.4$-point net difference, Cytokine Response-to-System changes 24.1\% for $-0.3$ points, and Drug Swapped-to-Matched changes 75.1\% for a $+13.0$-point gain. In each case, substantially more individual decisions change than the corresponding net accuracy difference suggests, because improvements can be offset by newly introduced errors or switches among incorrect answers. The positive panel-level Drug assignment effect also masks substantial model heterogeneity, revealing model-specific sensitivity to biological evidence assignment beyond aggregate accuracy. Additional diagnostics reveal sensitivity beyond aggregate accuracy: under Gene Base, all foundation models produce a single predicted category, while under Drug Matched, reversing candidate order changes 20.1\% of predictions despite preserving the biological target. Invalid outputs remain rare overall (187/59{,}760, 0.31\%). Together, these results show that endpoint accuracy alone can obscure how model decisions respond to controlled evidence interventions and task-preserving presentation changes.


\section{Conclusion}
We present \ours, a benchmark for evaluating foundation-model prediction behavior under controlled cross-scale biological evidence. Across \nmodel models, the highest accuracies are 57.5\% for Gene Dependency, 75.5\% for Cytokine Axis, and 70.8\% for Drug Sensitivity, leaving substantial predictive headroom. Evidence effects are selective: target-linked process annotations yield the largest Cytokine gains, matched reference histories outperform swapped histories on average in Drug Sensitivity, while additional transcriptional or system context provides little net benefit. Paired analyses further show that small accuracy differences can mask substantial prediction turnover. Overall, \ours enables fine-grained, task-specific attribution of prediction changes to controlled variations in evidence coverage, composition, and biological assignment without conflating these effects with underlying reasoning mechanisms.

\clearpage

\subsection*{AI Use Statement}

Generative AI tools, including OpenAI Codex, assisted with language editing, code development, and figure prototyping. They were not used as authoritative sources or to determine reference answers. The authors verified all AI-assisted text, citations, code, figures, and analyses and take full responsibility for the final work.

\subsection*{Ethics Statement}
We use existing cellular perturbation datasets and biological resources and conduct no new human or animal experiments.
Human-derived data are used only in their authorized, de-identified forms and in accordance with source licenses. \ours is intended for research evaluation and should not be used for clinical decision-making.

\subsection*{Reproducibility Statement}

The paper and appendix document the data resources, benchmark construction, prompts, model configurations, and evaluation procedures. Subject to source licenses and anonymous-review requirements, we will release the derived benchmark artifacts, construction and evaluation code, configurations, and detailed results.

\bibliography{iclr2027_conference}
\bibliographystyle{iclr2027_conference}

\appendix

\clearpage
\section{Related Work}
\label{sec:related_work}

\subsection{AI Virtual Cell Datasets and Models}

AI virtual cells seek to represent cellular states and predict their responses across perturbations and contexts~\citep{bunne2024virtualcell}. Large perturbation resources, including genome-scale Perturb-seq, scPerturb, and Tahoe-100M, provide experimentally grounded intervention--response measurements~\citep{replogle2022genotype,peidli2024scperturb,zhang2025tahoe}. Building on these data, single-cell foundation models learn transferable representations~\citep{theodoris2023geneformer,cui2024scgpt,hao2024scfoundation}, while CPA, CellOT, GEARS, STATE, and TxPert model perturbation-induced state transitions using latent, transport-based, graph-based, or set-level architectures~\citep{lotfollahi2023cpa,bunne2023cellot,roohani2024gears,adduri2026state,wenkel2026txpert}. Language-oriented approaches such as Cell2Sentence and VCWorld further connect cellular measurements with natural-language or knowledge-grounded reasoning~\citep{levine2024cell2sentence,wei2026vcworld}.

\subsection{Biological Reasoning Benchmarks for LLMs}

Recent benchmarks evaluate LLMs on increasingly biological forms of cellular reasoning. SOAR focuses on cell-type annotation~\citep{liu2025soar}; CellVerse and SC-Arena cover annotation, generation, perturbation prediction, and scientific question answering~\citep{zhang2025cellverse,zhao2026scarena}; and PerturbQA and VCWorld's GeneTAK evaluate differential-expression and direction prediction for unseen perturbations~\citep{wu2025perturbqa,wei2026vcworld}. BioMaze and PertReasonQA further assess pathway reasoning and the mechanistic faithfulness of perturbation explanations~\citep{zhao2025biomaze,kim2026pertreason}, while AssayBench extends evaluation to phenotypic CRISPR-screen prediction~\citep{debrouwer2026assaybench}.

\section{Experiment Setup}
\subsection{Configuration of Evaluated Models} \label{app:model_info}
\autoref{tab:model_configuration} details the configuration of each evaluated model. All inferences are conducted using either the official API endpoints or local \texttt{vLLM} deployment for most open-source models.
Our experiments are reproducible on a workstation equipped with two NVIDIA H200 GPUs.

\begingroup

\providecommand{\cmark}{}
\providecommand{\xmark}{}
\renewcommand{\cmark}{\ensuremath{\boldsymbol{\checkmark}}}
\renewcommand{\xmark}{\ensuremath{\boldsymbol{\times}}}

\setlength{\dashlinedash}{4pt}
\setlength{\dashlinegap}{3pt}
\setlength{\arrayrulewidth}{0.4pt}

\newcommand{\modeldashline}{%
  \noalign{\vskip 0.35ex}%
  \hdashline
  \noalign{\vskip 0.35ex}%
}

\begin{table*}[h]
\centering
\setlength{\tabcolsep}{5pt}
\renewcommand{\arraystretch}{1.08}

\resizebox{\textwidth}{!}{%
\begin{tabular}{@{}llllcc@{}}
\toprule

\textbf{Organization}
& \textbf{Model}
& \textbf{Release}
& \textbf{Version}
& \textbf{Reasoning?}
& \textbf{\makecell{Serving\\Backend}} \\

\midrule
\multicolumn{6}{c}{\emph{\textbf{Open-Source Models}}} \\
\midrule

\multirow{2}{*}{DeepSeek}
& DeepSeek-V4-Pro
& 2026-08
& \texttt{deepseek-ai/DeepSeek-V4-Pro-0813}
& \cmark
& vLLM \\

& DeepSeek-V4.1-Flash
& 2026-09
& \texttt{deepseek-ai/DeepSeek-V4.1-Flash}
& \cmark
& vLLM \\

\modeldashline

\multirow{2}{*}{Z.ai}
& GLM-5.2
& 2026-06
& \texttt{zai-org/GLM-5.2}
& \cmark
& vLLM \\

& GLM-5.3
& 2026-08
& \texttt{zai-org/GLM-5.3}
& \cmark
& vLLM \\

\modeldashline

MiniMax
& MiniMax-M3
& 2026-06
& \texttt{MiniMaxAI/MiniMax-M3}
& \cmark
& vLLM \\

\modeldashline

Moonshot AI
& Kimi-K3
& 2026-07
& \texttt{moonshotai/Kimi-K3}
& \cmark
& vLLM \\

\modeldashline

OpenAI
& GPT-OSS-20B
& 2025-08
& \texttt{openai/gpt-oss-20b}
& \cmark
& vLLM \\

\modeldashline

\multirow{2}{*}{Shanghai AI Lab}
& Agents-A1
& 2026-06
& \texttt{InternScience/Agents-A1}
& \cmark
& vLLM \\

& Intern-S1-mini
& 2025-08
& \texttt{internlm/Intern-S1-mini}
& \cmark
& vLLM \\

\modeldashline

\multirow{2}{*}{Tencent}
& Hunyuan-1.8B-Instruct
& 2025-07
& \texttt{tencent/Hunyuan-1.8B-Instruct}
& \cmark
& vLLM \\

& Hy3-FP8
& 2026-07
& \texttt{tencent/Hy3-FP8}
& \cmark
& vLLM \\

\modeldashline

\multirow{3}{*}{Qwen}
& Qwen3.6-35B-A3B
& 2026-04
& \texttt{Qwen/Qwen3.6-35B-A3B}
& \cmark
& vLLM \\

& Qwen3.8-27B
& 2026-08
& \texttt{Qwen/Qwen3.8-27B}
& \cmark
& vLLM \\

& Qwen3.8-Flash-Next
& 2026-08
& \texttt{Qwen/Qwen3.8-Flash-Next}
& \cmark
& vLLM \\

\modeldashline

\multirow{2}{*}{Meta}
& Llama-3.3-70B-Instruct
& 2024-12
& \texttt{meta-llama/Llama-3.3-70B-Instruct}
& \xmark
& vLLM \\

& Llama-4-Scout-17B-16E-Instruct
& 2025-04
& \texttt{meta-llama/Llama-4-Scout-17B-16E-Instruct}
& \xmark
& vLLM \\

\midrule
\multicolumn{6}{c}{\emph{\textbf{Proprietary Models}}} \\
\midrule

OpenAI
& GPT-5.6-Sol
& 2026-07
& \texttt{gpt-5.6-sol}
& \cmark
& API \\

\modeldashline

Google
& Gemini-3.8-Flash
& 2026-09
& \texttt{gemini-3.8-flash}
& \cmark
& API \\

\bottomrule
\end{tabular}%
}

\caption{Details of the foundation models evaluated in \ours.}
\label{tab:model_configuration}

\end{table*}
\endgroup

\subsection{Dataset Resources}

Table~\ref{tab:data_resources} summarizes the experimental datasets and external biological resources used in \ours, together with their data types and roles in benchmark construction.

\begin{table}[h]
\centering
\small
\caption{Inventory of the cross-scale data resources integrated in \ours.}
\label{tab:data_resources}

\resizebox{\linewidth}{!}{%
\begin{tabular}{@{}p{5cm}p{3.5cm}p{6cm}@{}}
\toprule
\textbf{Resource} &
\textbf{Data type} &
\textbf{Role in \ours} \\
\midrule

\multicolumn{3}{l}{\textbf{Core experimental datasets}} \\
\midrule

Tahoe-100M~\citep{zhang2025tahoe}
& Chemical perturbation
& Drug-induced single-cell transcriptional responses and perturbation conditions. \\

PBMC Parse10M~\citep{oesinghaus2025cytokine}
& Cytokine perturbation
& Cytokine-induced responses across immune cell types and donors. \\

Replogle Perturb-seq~\citep{replogle2022genotype}
& Genetic perturbation
& CRISPRi single-cell transcriptional responses in K562 and RPE1 cells. \\

\midrule
\multicolumn{3}{l}{\textbf{External functional endpoints and annotations}} \\
\midrule

PRISM~\citep{corsello2020discovering}
& Drug-response endpoint
& Drug--cell response measurements and viability-related sensitivity endpoints. \\

DepMap~\citep{tsherniak2017dependency}
& Genetic dependency
& Gene-effect and dependency measurements across cellular models. \\

PubChem~\citep{kim2023pubchem}
& Chemical annotation
& Compound identities, structures, and standardized chemical identifiers. \\

ChEMBL~\citep{zdrazil2024chembl}
& Bioactivity annotation
& Compound, target, assay, and bioactivity records. \\

BindingDB~\citep{gilson2016bindingdb}
& Binding evidence
& Protein--ligand binding affinities and interaction records. \\

UniProtKB~\citep{bateman2025uniprot}
& Protein annotation
& Protein identifiers, functions, and molecular annotations. \\

Reactome~\citep{milacic2024reactome}
& Pathway annotation
& Curated biological pathways and hierarchical pathway relations. \\

Gene Ontology and GOA~\citep{geneontology2023,huntley2015goa}
& Functional annotation
& Ontology terms and evidence-coded gene-function annotations. \\

Cellosaurus~\citep{bairoch2018cellosaurus}
& Cellular context
& Cell-line identities, aliases, cross-references, and background information. \\

scBaseCount~\citep{scbasecount}
& Sample and tissue context
& Standardized single-cell sample, tissue, and cellular-context metadata. \\

CORUM~\citep{tsitsiridis2023corum}
& Protein-complex annotation
& Protein-complex composition and subunit relations. \\

IntAct~\citep{deltoro2022intact}
& Molecular interaction
& Experimentally supported molecular interaction records. \\

SIGNOR~\citep{signor2023update}
& Signed mechanism relation
& Directed activation and inhibition relations for mechanistic reasoning. \\

Open Targets~\citep{ochoa2023opentargets}
& Target-association evidence
& Drug--target and target--disease associations with supporting evidence. \\

\midrule
\multicolumn{2}{r}{\textbf{Total source classes}} &
\textbf{17} \\

\bottomrule
\end{tabular}%
}
\end{table}

\clearpage

\clearpage

\section{Prompts}
\label{app:prompts}


We provide one representative instantiated prompt for each \ours QA type:
Gene Dependency Prediction in Figure ~\ref{fig:gene-dependency-prompt},
Cytokine-Axis Inverse Recovery in Figure ~\ref{fig:cytokine-axis-prompt},
and Drug-Sensitivity Ranking in Figure ~\ref{fig:drug-sensitivity-prompt}.

\begin{center}
\begin{minipage}{0.98\textwidth}

\begin{tcolorbox}[
    enhanced,
    colback=white,
    colframe=blue!45!black,
    boxrule=0.5pt,
    arc=2pt,
    left=6pt,
    right=6pt,
    top=4pt,
    bottom=5pt,
    title={Gene Dependency Prediction Prompt},
    colbacktitle=blue!12,
    coltitle=black,
    fonttitle=\normalsize\bfseries,
    fontupper=\small
]

\begin{tcolorbox}[
    colback=blue!4,
    colframe=blue!35,
    boxrule=0.4pt,
    arc=1.5pt,
    left=6pt,
    right=6pt,
    top=3pt,
    bottom=3pt,
    fontupper=\footnotesize
]
\textcolor{blue!55!black}{\textbf{\textsf{SYSTEM}}}\\[2pt]
You are a molecular and cellular biology expert. Evaluate the biological
question using only the supplied observations and visible evidence, and
select the best-supported answer from the candidates. Entity names,
accessions, aliases, and database identifiers may be masked. Do not
attempt to recover the hidden entity identity using external knowledge,
use hidden labels, or invent unavailable dependency measurements.
Return exactly one valid JSON object conforming to the required output
schema, without additional explanation or formatting.
\end{tcolorbox}

\vspace{-3pt}

\begin{tcolorbox}[
    colback=black!3,
    colframe=black!40,
    boxrule=0.4pt,
    arc=1.5pt,
    left=6pt,
    right=6pt,
    top=3pt,
    bottom=3pt,
    fontupper=\footnotesize
]
\textcolor{black!65}{\textbf{\textsf{USER}}}\\[2pt]

\textbf{Question}\\[1pt]
Infer the held-out gene-fitness dependency endpoint category for an
anonymous CRISPRi perturbation in K562.

\vspace{3pt}
\textbf{Task Requirements}\\[1pt]
General protein annotations and pathway membership provide indirect
biological evidence and should not be treated as measurements of the
dependency endpoint. Infer the endpoint only from the evidence visible
under the current exposure condition.

\vspace{3pt}
\textbf{Evidence Exposure Condition}\\[1pt]
\texttt{Pathway context}

\vspace{3pt}
\textbf{Query Context}\\[1pt]
\textbf{Cell line:} \texttt{K562}.\\
\textbf{Perturbation type:} \texttt{CRISPRi}.\\
\textbf{Perturbation entity:}
\texttt{ANON\_PERTURBATION\_GENE}.

\vspace{3pt}
\textbf{Visible Evidence}\\[1pt]
The anonymous perturbation target is associated with the following
Reactome pathway annotations: Signaling by EGFR, Signaling by Receptor
Tyrosine Kinases, Signaling by VEGF, Signal Amplification, NOD1/2
Signaling, NLR Signaling, Hemostasis, and Immune System. Explicit gene
names, protein names, accessions, aliases, and database identifiers are
withheld.

\vspace{3pt}
\textbf{Candidate Answers}\\[1pt]
\textbf{A:} A strong-dependency endpoint category with a substantially
negative CRISPR gene-effect value.\\
\textbf{B:} A moderate-dependency endpoint category with an intermediate
negative CRISPR gene-effect value.\\
\textbf{C:} A weak-or-no-dependency endpoint category with a gene-effect
value near the nonessential range.

\vspace{3pt}
\textbf{Required Output}\\[1pt]
\texttt{\{}\\
\hspace*{1em}\texttt{"evidence\_sufficiency":
"sufficient|weak|insufficient",}\\
\hspace*{1em}\texttt{"best\_effort\_dependency\_label":}\\
\hspace*{2em}\texttt{"strong\_dependency|moderate\_dependency|}\\
\hspace*{2em}\texttt{weak\_or\_no\_dependency",}\\
\hspace*{1em}\texttt{"effect\_direction":}\\
\hspace*{2em}\texttt{"stronger\_dependency|weaker\_dependency|}\\
\hspace*{2em}\texttt{neutral\_or\_unknown",}\\
\hspace*{1em}\texttt{"confidence": "low|medium|high"}\\
\texttt{\}}

\end{tcolorbox}
\end{tcolorbox}

\captionof{figure}{
Representative instantiated prompt for Gene Dependency Prediction under
the pathway-context condition. Direct entity identifiers and database
cross-references are masked, while task-relevant functional annotations
remain visible. The dependency endpoint and reference answer are
withheld from the model; evidence records are abbreviated for readability.
}
\label{fig:gene-dependency-prompt}

\end{minipage}
\end{center}

\begin{center}
\begin{minipage}{0.98\textwidth}
\begin{tcolorbox}[
    enhanced,
    colback=white, colframe=green!40!black,
        boxrule=0.5pt, arc=2pt,
        left=6pt, right=6pt, top=4pt, bottom=5pt,
        title={Cytokine-Axis Inverse Recovery Prompt},
    colbacktitle=green!10, coltitle=black,
        fonttitle=\normalsize\bfseries, fontupper=\small
    ]
    
    \begin{tcolorbox}[
        colback=blue!4, colframe=blue!35,
        boxrule=0.4pt, arc=1.5pt,
        left=6pt, right=6pt, top=3pt, bottom=3pt,
        fontupper=\footnotesize
    ]
    \textcolor{blue!55!black}{\textbf{\textsf{SYSTEM}}}\\[2pt]
    You are a molecular and cellular biology expert. Evaluate the biological
    question using the supplied observations and context, and select the
    best-supported answer from the candidates. The queried perturbation may be
    anonymized. Use only the information provided in the input and do not infer
    information that is absent. Return exactly one valid JSON object without
    additional explanation or formatting.
    \end{tcolorbox}
    
    \vspace{-3pt}
    
    \begin{tcolorbox}[
        colback=black!3, colframe=black!40,
        boxrule=0.4pt, arc=1.5pt,
        left=6pt, right=6pt, top=3pt, bottom=3pt,
        fontupper=\footnotesize
    ]
    \textcolor{black!65}{\textbf{\textsf{USER}}}\\[2pt]
    
    \vspace{3pt}
    \textbf{Question}\\[1pt]
    Predict the held-out cytokine ligand/receptor axis for an anonymous cytokine
    stimulus in \texttt{CD14 Mono} cells.
    
    \vspace{3pt}
    \textbf{Task Requirements}\\[1pt]
    Use the measured single-cell response together with the visible
    biological-process and system-context evidence.
    
    \vspace{3pt}
    \textbf{Observations and Context}\\[1pt]
    \textbf{Measured response:} relative to matched PBS controls, representative
    top-down genes are \texttt{GBP1} ($-0.371531$), \texttt{CXCL10}
    ($-0.249265$), \texttt{MARCKS} ($-0.246084$), \texttt{CD163}
    ($-0.236194$), and \texttt{ANKRD22} ($-0.219791$), \ldots;
    representative top-up genes are \texttt{PLXDC2} ($0.304950$), \texttt{CD68}
    ($0.257540$), \texttt{SIGLEC1} ($0.250829$), \texttt{CD74}
    ($0.244919$), and \texttt{ACTB} ($0.208748$), \ldots.\\
    \textbf{Biological-process evidence:} helper T-cell enhancement of adaptive
    immune response; leukocyte chemotaxis; induction of positive chemotaxis;
    chemotaxis; T-cell activation; regulation of T-cell activation; positive
    regulation of T-cell proliferation; and T-cell receptor signaling.\\
    \textbf{System context:} \texttt{CD14 Mono} cells from donors
    \texttt{Donor1}, \texttt{Donor3}, \texttt{Donor5}, \texttt{Donor6},
    \texttt{Donor8}, \texttt{Donor10}, and \texttt{Donor12}.
    
    \vspace{3pt}
    \textbf{Candidate Answers}\\[1pt]
    \textbf{A:} ligand genes = \texttt{LEP}; receptor genes = \texttt{LEPR}.\\
    \textbf{B:} ligand genes = \texttt{IFNW1}; receptor genes =
    \texttt{IFNAR1, IFNAR2}.\\
    \textbf{C:} ligand genes = \texttt{IL1A}; receptor genes =
    \texttt{IL1R1, IL1RAP}.\\
    \textbf{D:} ligand genes = \texttt{IL16}; receptor genes = \texttt{CD4}.
    
    \vspace{3pt}
    \textbf{Required Output}\\[1pt]
    \texttt{\{"selected\_answer\_label": "A|B|C|D"\}}
    \end{tcolorbox}
    
\end{tcolorbox}
\captionof{figure}{Representative instantiated prompt for Cytokine-Axis Inverse Recovery;
repeated evidence records are abbreviated for readability.}
\label{fig:cytokine-axis-prompt}
\end{minipage}
\end{center}

\begin{center}
\begin{minipage}{0.98\textwidth}
\begin{tcolorbox}[
    enhanced,
    colback=white,
    colframe=magenta!40!black,
    boxrule=0.5pt,
    arc=2pt,
    left=6pt,
    right=6pt,
    top=4pt,
    bottom=5pt,
    title={Drug-Sensitivity Ranking Prompt},
    colbacktitle=magenta!10,
    coltitle=black,
    fonttitle=\normalsize\bfseries,
    fontupper=\small
]

\begin{tcolorbox}[
    colback=blue!4,
    colframe=blue!35,
    boxrule=0.4pt,
    arc=1.5pt,
    left=6pt,
    right=6pt,
    top=3pt,
    bottom=3pt,
    fontupper=\footnotesize
]
\textcolor{blue!55!black}{\textbf{\textsf{SYSTEM}}}\\[2pt]

You are a molecular and cellular biology expert. Evaluate the biological question using the supplied observations and visible evidence, and select the best-supported answer from the candidates. The queried treatment and cell-model identities may be anonymized. Use only the information provided in the input and do not infer unavailable endpoint values. Return exactly one valid JSON object without additional explanation or formatting.
\end{tcolorbox}

\vspace{-3pt}

\begin{tcolorbox}[
    colback=black!3,
    colframe=black!40,
    boxrule=0.4pt,
    arc=1.5pt,
    left=6pt,
    right=6pt,
    top=3pt,
    bottom=3pt,
    fontupper=\footnotesize
]
\textcolor{black!65}{\textbf{\textsf{USER}}}\\[2pt]

\vspace{3pt}
\textbf{Question}\\[1pt]

Which candidate cell-line multidose response trajectory is more sensitive to the fixed treatment stimulus \texttt{ANON\_STIMULUS}?

\vspace{3pt}
\textbf{Task Requirements}\\[1pt]

Use the balanced response summaries and reference-history evidence. Lower held-out external AUC indicates stronger sensitivity. The held-out query AUC is not used to construct the reference histories.

\vspace{3pt}
\textbf{Observations and Context}\\[1pt]

\textbf{Dose response:} At $[0.05, 0.5, 5.0]$ $\mu$M, mean absolute response trajectories are $[0.29015, 0.27761, 0.45911]$ for A and $[0.48819, 0.38289, 0.96108]$ for B.\\

\textbf{Gene extremes:} At high dose, A includes down-regulated \texttt{7SK}, \texttt{MTND2P28}, and \texttt{RNF26}, and up-regulated \texttt{ASMTL-AS1}, \texttt{RBMS3-AS3}, and \texttt{YBX1P10}; B includes down-regulated \texttt{ABHD6}, \texttt{MTND2P28}, and \texttt{TCIRG1}, and up-regulated \texttt{AFP}, \texttt{LBH}, and \texttt{USP38-DT}.\\

\textbf{Module summaries:} High-dose and low-to-high signed responses are provided for apoptosis/stress, cell cycle, DNA damage/p53, interferon, MAPK/ERK, NF-$\kappa$B, PI3K/AKT/mTOR, and senescence/SASP modules.\\

\textbf{Reference histories:} Development-set sensitivity percentiles associated with the corresponding candidate cell models include $0.583333$ for A and $0.062500$, $0.062500$, and $0.055556$ for B, \ldots. Larger percentiles indicate greater relative sensitivity within the corresponding reference compound.\\

\textbf{Target evidence:} BindingDB records link the anonymous ligand to \texttt{PLK1} (IC$_{50}$ $=0.87$--$2.4$ nM) and \texttt{BRD4} (K$_d$ $=77$--$255$ nM; IC$_{50}=1070$ nM), with additional UniProt context for \texttt{BRD4}.

\vspace{3pt}
\textbf{Candidate Answers}\\[1pt]

\textbf{A:} First anonymous cell-line multidose response trajectory.\\
\textbf{B:} Second anonymous cell-line multidose response trajectory.

\vspace{3pt}
\textbf{Required Output}\\[1pt]

\texttt{\{"preferred\_answer\_label": "A|B"\}}

\end{tcolorbox}

\end{tcolorbox}

\captionof{figure}{Representative instantiated prompt for Drug-Sensitivity Ranking under the matched-reference-history condition; repeated evidence records are abbreviated for readability.}
\label{fig:drug-sensitivity-prompt}

\end{minipage}
\end{center}

\clearpage


\section{Evidence Attribution}
\label{app:evidence_attribution}

The controlled-evidence analysis relies on task-specific evidence constructions and paired evaluation conditions. We therefore report additional details on case eligibility, benchmark selection, and evidence construction, together with the condition-level measurements underlying the contrasts in the main text.

\paragraph{Case Eligibility and Benchmark Selection.}
Across tasks, a case is considered eligible only when its biological entity and experimental context can be aligned, the required task-specific measurements and evidence are available, and the reference target maps to a unique candidate answer. The question, candidate set, reference target, and scoring rule are then held fixed across matched evidence conditions, with prohibited query-endpoint fields excluded from model-visible inputs. Failure of any applicable criterion excludes the corresponding case and all of its matched conditions.

Source-record and candidate-case counts use different units and therefore do not define a direct attrition sequence. For Gene Dependency, 2,448 endpoint-labeled candidates entered the evidence-availability and boundary-margin checks, of which 2,092 satisfied the eligibility criteria. We then sampled 40 cases from each dependency-category--cell-line stratum to construct the 240-case benchmark set. For Cytokine Axis, 267 cases satisfied the task-specific eligibility criteria, from which 110 cases were selected with ten cases per target axis. For Drug Sensitivity, 737 candidate pairs satisfied the curve-quality, complete-dose-trajectory, and AUC-separation criteria; 240 pairs were selected for benchmark evaluation to increase compound coverage while limiting reuse of drug--cell-model units. Differences between eligible and benchmark case counts therefore reflect benchmark-set construction rather than additional case-level quality-control failures. The frozen construction artifacts do not provide mutually exclusive counts for every earlier rejection criterion, so we do not further partition the remaining source records by failure reason.

\paragraph{Cytokine Process Evidence.}
The Cytokine conditions retain the measured transcriptional response while varying the additional process and system context. Process annotations are linked to the recorded cytokine stimulus, whereas system evidence describes the corresponding cellular context. The Process--Response contrast therefore measures the predictive value associated with the target-linked process context supplied to the model. Response--mechanism integration is not separately identified by this contrast. The Joint condition combines Process and System evidence under the same overall evidence budget by reallocating the available evidence units between the two sources. Accordingly,
\[
\Gamma
=
A_{\mathrm{Joint}}
-
A_{\mathrm{Process}}
-
A_{\mathrm{System}}
+
A_{\mathrm{Response}}
\]
is treated as a joint-condition deviation from additivity under this budget constraint.

\paragraph{Drug Reference-History Construction.}
Drug reference histories are constructed from 36 development compounds that are disjoint from the query compounds. Development records are retained after PRISM quality control requiring median fit $R^2 \geq 0.5$, replicate AUC range $\leq 0.1$, positive slope, and ordered asymptotes. Within each development compound, relative sensitivity is represented by within-compound AUC-based percentiles for the corresponding cell models. Query-compound AUC values and query transcriptional responses are not used to construct or select these reference-history records.

The Matched and Swapped conditions contain the same query-response information and use the same history-comparison instruction. In the Matched condition, each reference history remains associated with the cell model from which it was derived; in the Swapped condition, the histories are exchanged between the two candidates. The Matched--Swapped contrast therefore measures sensitivity to candidate--history assignment conditional on the supplied query and reference-history evidence.

\paragraph{Drug Reference-History Signal Diagnostic.}
After selecting the 240-case Drug benchmark cohort, we evaluated whether its development-set reference histories contained predictive signal using a simple evidence-only rule. For each candidate cell model, the rule averages its sensitivity percentiles across available development-compound histories and selects the candidate with the higher mean. Predictions are scored against the held-out query label, and accuracy is averaged equally across the 26 query compounds. The rule attains 64.2\% compound-weighted accuracy, exceeding the prespecified 60\% cohort-level threshold.

The diagnostic is applied to the fixed benchmark cohort and does not remove individual cases. Because its evaluation uses held-out query labels after benchmark-set selection, we interpret the resulting accuracy as a post-selection measure of reference-history informativeness. The Drug evidence contrasts reported in the main text therefore characterize the benchmark cohort rather than all 737 eligible candidate pairs.

\paragraph{Condition-Level Measurements.}
Table~\ref{tab:condition_accuracy} reports biological-cluster-weighted accuracy under each evidence condition. These values provide the condition-level measurements underlying the controlled-evidence contrasts in Table~\ref{tab:main_results}b. Gene Dependency contrasts weight perturbed genes equally, Cytokine Axis contrasts weight cytokine axes equally, and Drug Sensitivity contrasts weight query compounds equally. For Drug Sensitivity, both candidate-order presentations remain paired within each case, and invalid outputs receive zero score under the fixed denominator.


\newcommand{\rowstretch}{1.25}   
\newcommand{\vlh}{10.5pt}        
\newcommand{\vld}{4.5pt}         
\newcolumntype{V}{!{\vrule width \arrayrulewidth height \vlh depth \vld}}

\begin{table*}[h]
\centering
\caption{Biological-cluster-weighted accuracy across four evidence conditions per task (\%). Differences of unrounded condition values reproduce Table~\ref{tab:main_results}b; displayed values may differ slightly after rounding. This weighting differs from the case-weighted endpoint accuracies in Table~\ref{tab:main_results}a. Drug retains both candidate orderings, and invalid outputs score zero.}
\label{tab:condition_accuracy}
\setlength{\tabcolsep}{5pt}
\renewcommand{\arraystretch}{\rowstretch}
\setlength{\aboverulesep}{0pt}
\setlength{\belowrulesep}{0pt}

\resizebox{\textwidth}{!}{%
\begin{tabular}{@{}lVrrrrVrrrrVrrrr@{}}
\toprule
Model & \multicolumn{4}{cV}{Gene Dependency} & \multicolumn{4}{cV}{Cytokine Axis} & \multicolumn{4}{c@{}}{Drug Sensitivity} \\
& Base & Mol. & Path. & Trans. & Resp. & Proc. & System & Joint & Back. & Query & Matched & Swapped \\
\midrule
\multicolumn{13}{@{}>{\columncolor{gray!15}}c@{}}{\textit{Open-Source Models}} \\
\midrule
DeepSeek-V4-Pro & 36.1 & 36.5 & 44.7 & 43.6 & 23.6 & 65.5 & 24.5 & 66.4 & 50.8 & 52.9 & 63.4 & 36.5 \\
GLM-5.2 & 36.1 & 36.1 & 47.9 & 47.5 & 15.5 & 60.0 & 15.5 & 53.6 & 49.4 & 42.6 & 58.4 & 38.6 \\
MiniMax-M3 & 36.1 & 36.5 & 46.8 & 45.0 & 18.2 & 61.8 & 11.8 & 65.5 & 49.6 & 43.4 & 60.7 & 35.7 \\
Kimi-K3 & 36.1 & 53.2 & 49.8 & 50.9 & 17.3 & 78.2 & 24.5 & 70.0 & 41.5 & 41.8 & 63.6 & 36.6 \\
DeepSeek-V4.1-Flash & 36.1 & 41.8 & 52.1 & 51.4 & 30.0 & 69.1 & 33.6 & 60.9 & 50.1 & 40.5 & 42.7 & 48.0 \\
GLM-5.3 & 36.1 & 49.5 & 53.0 & 53.4 & 20.0 & 74.5 & 24.5 & 70.9 & 49.9 & 44.4 & 60.7 & 34.8 \\
Qwen3.6-35B-A3B & 36.1 & 36.1 & 36.5 & 37.0 & 15.5 & 61.8 & 13.6 & 60.0 & 50.0 & 53.8 & 62.1 & 38.2 \\
Qwen3.8-27B & 36.1 & 36.1 & 36.1 & 36.8 & 10.9 & 64.5 & 10.0 & 59.1 & 47.9 & 46.9 & 64.0 & 35.8 \\
GPT-OSS-20B & 36.1 & 36.1 & 36.1 & 36.1 & 19.1 & 40.9 & 14.5 & 38.2 & 49.4 & 51.2 & 64.2 & 35.8 \\
Agents-A1 & 36.1 & 36.1 & 37.4 & 37.7 & 17.3 & 58.2 & 14.5 & 45.5 & 50.0 & 47.2 & 47.6 & 48.5 \\
Intern-S1-mini & 36.1 & 36.1 & 36.5 & 35.6 & 7.3 & 48.2 & 8.2 & 40.0 & 50.0 & 46.0 & 43.2 & 55.0 \\
Hunyuan-1.8B-Instruct & 28.3 & 28.3 & 28.8 & 27.6 & 24.5 & 24.5 & 24.5 & 25.5 & 50.0 & 50.0 & 50.0 & 50.0 \\
Qwen3.8-Flash-Next & 36.1 & 36.1 & 51.6 & 46.6 & 25.5 & 81.8 & 20.9 & 74.5 & 50.0 & 47.6 & 62.2 & 36.1 \\
Llama-3.3-70B-Instruct & 36.1 & 36.1 & 36.1 & 36.1 & 11.8 & 39.1 & 9.1 & 32.7 & 50.0 & 50.1 & 39.2 & 59.8 \\
Llama-4-Scout-17B-16E-Instruct & 36.1 & 36.1 & 36.1 & 36.1 & 16.4 & 48.2 & 15.5 & 36.4 & 50.0 & 50.3 & 42.4 & 58.9 \\
Hy3-FP8 & 36.1 & 36.1 & 36.1 & 36.1 & 10.9 & 68.2 & 11.8 & 60.9 & 50.0 & 42.1 & 50.7 & 45.7 \\
\midrule
\multicolumn{13}{@{}>{\columncolor{gray!15}}c@{}}{\textit{Proprietary Models}} \\
\midrule
GPT-5.6-Sol & 36.1 & 40.9 & 54.3 & 54.3 & 28.2 & 79.1 & 27.3 & 71.8 & 50.2 & 41.3 & 60.8 & 35.1 \\
Gemini-3.8-Flash & 36.1 & 54.1 & 49.8 & 50.2 & 30.0 & 80.9 & 31.8 & 75.5 & 50.0 & 42.8 & 63.7 & 37.2 \\
\bottomrule
\end{tabular}%
}
\end{table*}

\section{Biological-Cluster Bootstrap Confidence Intervals}
\label{app:bootstrap_CI}

We estimate uncertainty using 10,000 biological-cluster bootstrap replicates with the task-specific clusters defined in Section~\ref{sec:evaluation_metrics}. In each replicate, clusters are sampled with replacement while retaining all associated cases, candidate-order presentations, evidence conditions, and model predictions jointly. For Drug Sensitivity, both candidate-order presentations of each case remain within the same compound cluster. Endpoint accuracy averages presentation scores within each case and weights cases equally, whereas controlled contrasts weight biological clusters equally. We report percentile-based 95\% confidence intervals using the 2.5th and 97.5th percentiles of the bootstrap distribution.

Table~\ref{tab:model_inference_ci} reports model-level case-weighted endpoint accuracies and their 95\% confidence intervals under the prespecified target-evidence conditions. Table~\ref{tab:evidence_effect_inference} reports the corresponding aggregate controlled-evidence effects across the 18-model panel, including incremental, joint-deviation, and assignment contrasts. The latter averages paired case-score differences within biological clusters and then averages the resulting model-level effects across the fixed model panel.

These intervals are marginal under a single perturbation-clustering scheme. Shared donors, controls, cell models, and reference histories may induce dependence across clusters, and the intervals do not establish robustness to such crossed dependence. The model panel is fixed, so they do not quantify uncertainty over model families or future models. Intervals containing zero should not be interpreted as evidence of equivalence.



\begin{table*}[h]
  \centering
  \caption{Case-weighted endpoint accuracy with 95\% biological-cluster bootstrap confidence intervals for the formal 18-model panel (\%).}
  \label{tab:model_inference_ci}
  \footnotesize
  \setlength{\tabcolsep}{5pt}
  \renewcommand{\arraystretch}{1.08}
  \setlength{\aboverulesep}{0pt}
  \setlength{\belowrulesep}{0pt}
  \resizebox{0.82\textwidth}{!}{%
  \begin{tabular}{@{}lccc@{}}
    \toprule
    \textbf{Model} & \textbf{Gene Dependency} & \textbf{Cytokine Axis} & \textbf{Drug Sensitivity} \\
    \midrule
    \multicolumn{4}{c}{\cellcolor{gray!15}\textit{Open-Source Models}} \\
    \midrule
    DeepSeek-V4-Pro & 42.1 [35.6, 48.7] & 66.4 [42.7, 87.3] & 69.8 [61.4, 77.5] \\
    GLM-5.2 & 46.7 [40.0, 53.4] & 53.6 [30.0, 76.4] & 62.9 [54.2, 71.4] \\
    MiniMax-M3 & 44.2 [37.7, 50.8] & 65.5 [45.5, 83.6] & 67.7 [59.6, 75.1] \\
    Kimi-K3 & 54.2 [47.4, 60.8] & 70.0 [47.3, 89.1] & 70.0 [61.6, 77.7] \\
    DeepSeek-V4.1-Flash & 52.9 [46.4, 59.5] & 60.9 [39.1, 81.8] & 43.8 [37.5, 50.2] \\
    GLM-5.3 & 55.0 [48.1, 61.8] & 70.9 [50.0, 89.1] & 68.5 [59.9, 76.5] \\
    Qwen3.6-35B-A3B & 34.6 [28.4, 40.9] & 60.0 [36.4, 81.8] & 67.9 [60.1, 75.2] \\
    Qwen3.8-27B & 34.2 [28.0, 40.5] & 59.1 [36.4, 80.9] & 70.6 [62.4, 78.1] \\
    GPT-OSS-20B & 33.3 [27.2, 39.7] & 38.2 [22.7, 53.6] & 70.8 [62.7, 78.2] \\
    Agents-A1 & 35.0 [28.9, 41.4] & 45.5 [22.7, 68.2] & 46.0 [38.8, 53.8] \\
    Intern-S1-mini & 32.9 [26.9, 39.3] & 40.0 [20.9, 59.1] & 40.4 [35.2, 45.9] \\
    Hunyuan-1.8B-Instruct & 32.5 [25.8, 39.3] & 25.5 [19.1, 32.7] & 50.0 [50.0, 50.0] \\
    Qwen3.8-Flash-Next & 47.9 [41.3, 54.6] & 74.5 [53.6, 90.9] & 69.6 [61.3, 77.1] \\
    Llama-3.3-70B-Instruct & 33.3 [27.2, 39.7] & 32.7 [19.1, 48.2] & 33.3 [26.7, 40.6] \\
    Llama-4-Scout-17B-16E-Instruct & 33.3 [27.2, 39.7] & 36.4 [21.8, 50.0] & 35.4 [28.0, 43.7] \\
    Hy3-FP8 & 33.3 [27.2, 39.7] & 60.9 [40.9, 80.0] & 55.4 [51.6, 59.0] \\
    \midrule
    \multicolumn{4}{c}{\cellcolor{gray!15}\textit{Proprietary Models}} \\
    \midrule
    GPT-5.6-Sol & 57.5 [51.0, 64.0] & 71.8 [50.9, 90.0] & 68.8 [60.3, 76.3] \\
    Gemini-3.8-Flash & 53.3 [46.6, 60.0] & 75.5 [50.9, 94.5] & 70.2 [61.7, 78.0] \\
    \bottomrule
  \end{tabular}%
  }
\end{table*}

\begin{table}[h]
  \centering
  \caption{Aggregate controlled-evidence effects across the formal 18-model panel (percentage points).}
  \label{tab:evidence_effect_inference}
  \footnotesize
  \setlength{\tabcolsep}{5pt}
  \renewcommand{\arraystretch}{1.08}
  \begin{tabular}{@{}llcc@{}}
    \toprule
    \textbf{Task} & \textbf{Effect} & \textbf{Aggregate $\Delta_{\mathrm{Acc}}$} & \textbf{95\% CI} \\
    \midrule
    \multirow{3}{*}{Gene Dependency} & Molecular $-$ Base & +3.3 & [+2.2, +4.5] \\
     & Pathway $-$ Molecular & +3.8 & [+1.7, +5.9] \\
     & Transcriptional $-$ Pathway & $-$0.4 & [$-$1.1, +0.3] \\
    \midrule
    \multirow{3}{*}{Cytokine Axis} & Process $-$ Response & +42.4 & [+28.2, +56.0] \\
     & System $-$ Response & $-$0.3 & [$-$2.0, +1.4] \\
     & Joint deviation & $-$5.1 & [$-$8.9, $-$1.6] \\
    \midrule
    \multirow{3}{*}{Drug Sensitivity} & Query Response $-$ Background & $-$3.0 & [$-$6.6, +1.0] \\
     & Matched $-$ Background & +6.2 & [$-$0.3, +12.2] \\
     & Matched $-$ Swapped & +13.0 & [+0.5, +24.5] \\
    \bottomrule
  \end{tabular}
\end{table}


\end{document}